\documentclass[twocolumn,showpacs,preprintnumbers,amsmath,amssymb]{revtex4}
\usepackage{epsfig}
\usepackage{amsmath}

\begin{document}

\title{On the Diversity of Non-Linear Transient Dynamics in Several 
Types of Complex Networks}

\author{Luciano da Fontoura Costa}
\affiliation{Institute of Physics at S\~ao Carlos, University of
S\~ao Paulo, PO Box 369, S\~ao Carlos, S\~ao Paulo, 13560-970 Brazil}

\date{15th Dec 2007}

\begin{abstract}
Dynamic systems characterized by diversified evolutions are not only
more flexible, but also more resilient to attacks, failures and
changing conditions.  This article addresses the quantification of the
diversity of non-linear transient dynamics obtained in undirected and
unweighted complex networks as a consequence of self-avoiding random
walks.  The diversity of walks starting at a specific node $i$ is
quantified in terms of a signature composed by the entropies of the
node visit probabilities along each of the initial steps.  Six
theoretical models of complex networks are considered:
Erd\H{o}s-R\'enyi, Barab\'asi-Albert, Watts-Strogatz, a geographical
model, as well as two recently introduced knitted networks formed by
paths.  The random walk diversity is explored at the level of network
categories and of individual nodes.  Because the diversity at
successive steps of the walks tends to be correlated, principal
component analysis is systematically applied in order to identify the
more relevant linear combinations of the diversity entropies and to
obtain optimal dimensionality reduction.  Several interesting results
are reported, including the facts that the transient diversity tends
to increase with the average degree for all considered network models
and that the Watts and Strogatz and geographical models tend to yield
diversity entropies which increase more gradually with the number of
steps, contrasting sharply with the steep increases verified for the
other four considered models.  The principal linear combination of the
diversities identified by the principal component analysis method is
shown to allow an interesting characterization of individual nodes as
well as partitioning of networks into subgraphs of similar diversity.
\end{abstract}

\maketitle

\vspace{0.5cm}
\emph{`There is a city where you arrive for the first time;
and there is another city which you leave never to return.' 
(Invisible Cities, I. Calvino)}

\section{Introduction} 

The diversity of dynamics plays a key role in most aspects of nature,
which has ultimately resulted in a wealthy of species along evolution
as well as a myriad of human cultural manifestations.  Because of
their capacity to represent discrete structures and scaffold dynamics,
complex networks (e.g.~\cite{Albert_Barab:2002, Dorogov_Mendes:2002,
Newman:2003, Boccaletti:2006, Costa_surv:2007}) have become the key
paradigm in theoretical and applied studies in complex dynamic
systems, finding applications in an impressive range of problems
(e.g.~\cite{Costa_surv:2007}).  A great deal of the current attention
in this area concentrates not only in characterizing the topological
properties of networks (e.g.~\cite{Costa_surv:2007}, but also in
investigating how the latter constrains or even define dynamics
unfolding in the networks (e.g.~\cite{Newman:2003, Boccaletti:2006}).

With a tradition extending back over several decades, the study of the
dynamics of random walks represents one of the main paradigms in
statistical physics and dynamical systems.  Traditional random walks
are usually performed by one or more agents choosing with uniform
probability between the outgoing edges at each node.  Therefore,
random walks represent one of the least intelligent ways to move in a
network, involving no additional criterion rather than uniform chance.
Still, such a dynamics is directly related to the important linear
dynamics of diffusion (e.g.~\cite{Doyle_Snell:1984, Sethna:2006}),
which plays an important role in a large number of natural dynamical
processes (e.g. reaction-diffusion and Schr\"odinger equation).  The
dynamics of traditional, linear, random walks on complex networks has
been investigated by several articles
(e.g.~\cite{Zhou:2003,Rieger:2004, Masuda:2004, Pons_comm:2005,
Eisler:2005, Costa_corrs:2007}).  Several other types of random walks
have also been considered in the literature
(e.g.~\cite{Kinouchi_thesaurus:2001,Yang:2005, Costa_know:2006}).  For
instance, the category of \emph{self-avoiding} random walks represents
a particularly interesting situation in which the moving agent is not
allowed to return to nodes and/or edges.  As such, self-avoiding walks
are can be directly associated to the paths existing in the networks.
By \emph{path}, it is henceforth meant a sequence of
adjacent~\footnote{Two undirected edges $(i,j)$ and $(p,q)$ are
adjacent iff $i=p$ or $i=q$ or $j=p$ or $j=q$.} edges without
repetition of node or edge.  Paths are important because they provide
the most effective way to connect the involved nodes (i.e. given $M$
nodes, a path through them involves $M-1$ edges).  In addition, unlike
random walks, random self-avoiding walks ---
\emph{path-walks} for short --- are non-linear and necessarily finite 
in finite networks, because the moving agent sooner or later has no
way to proceed.  The possibility to use self-avoiding walks to sample
networks has been investigated in~\cite{Yang:2005}.  Paths and
self-avoiding walks have recently been explored as dual motifs of star
connectivity~\cite{Costa_path:2007}, building block of
networks~\cite{Costa_comp:2007} and for characterization of networks
(especially through the longest path)~\cite{Costa_longest:2007}.  The
transient dynamics of self-avoiding walks in uniformly random, small
world and scale free networks has been studied
in~\cite{Herrero_self:2003, Herrero_self:2005}, with special attention
placed on the average number of such walks.

\begin{figure}
  \vspace{0.3cm}
  \begin{center}
  \includegraphics[width=0.9\linewidth]{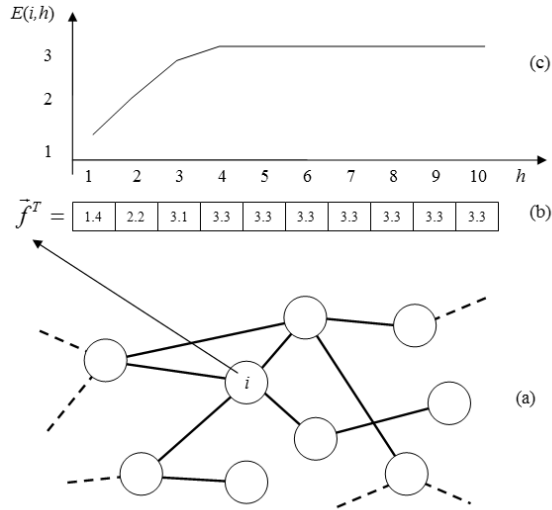}
  \caption{Given a node $i$ in a network (a), the entropies $E(i,h)$
  of the probabilities of visited nodes after the initial $h$ steps
  (c) can be calculated by simulating several self-avoiding random walks
  starting from $i$. Therefore, a signature $\vec{f}$ (b)can assigned
  to each node $i$ which expresses the diversity of paths obtained
  at each step $h$.}~\label{fig:features}
  \end{center}
\end{figure}

Random walks typically start from a node and proceed~\footnote{It is
sometimes interesting~\cite{Costa_comp:2007,Costa_longest:2007} to
proceed in two directions, i.e. through two outgoing edges from the
starting node. Though interesting, this type of self-avoiding walk is
not considered in this work.}  until some stopping condition is met
(e.g. fixed number of steps or, in the case of self-avoiding walks,
impossibility to proceed further).  In this work all walks are
performed for a pre-specified number of 10 steps, as we are interested
in the transient dynamics.  Consider now that the starting node has
been fixed and several self-avoiding random walks are performed from
that node.  One interesting question regards how such walks are
composed and distributed.  For instance, one may be interested in the
length of these walks
(e.g.~\cite{Herrero_self:2005,Costa_longest:2007}).  A question of
special relevance which has received little attention from the
literature regards the \emph{diversity} of the obtained walks and
path-walks.  By diversity, it is meant how much the walks differ one
another by incorporating distinct nodes and/or edges.  In a previous
approach to this problem, Herrero investigated the average number of
self-avoiding walks defined in uniformly random and scale free
networks.  In the present article, the diversity of self-avoiding
walks starting at a specific node $i$ is quantified in terms of the
entropies of the node probability visits after the first $S$ steps
along the walk after starting from each node $i$, giving rise to
diversity entropy (see Figure~\ref{fig:features}).  Because of the
non-linear nature of this type of walks and our interest in obtaining
information about each individual walk in several types of
structurally diverse networks, the visit probabilities are estimated
by performing several self-avoiding walks.

Provided such a diversity of dynamics can be quantified, ideally in
terms of a single measurement, a series of interesting analyses can be
performed at several levels (e.g. from individual node to network
category levels).  Indeed, the diversity of walks is immediately
related to a large number of important theoretical and practical
aspects of complex networks structure and dynamics.  To begin with,
the cases in which the path-walks are found to be mostly similar
(i.e. little diversity) imply that the agent had little choice during
its motion, and therefore little path redundancy is present in the
network, starting from that node.  At the same time, such situations
will also be characterized as being highly efficient as far as node
coverage is concerned (relatively few edges are required while
visiting several nodes).  Indeed, by recalling that all nodes in a
path-walk must be distinct, a path-walk involving $S+1$ nodes will
necessarily have $S$ edges, which is the minimum number of connections
required to connected those nodes.  On the contrary, in case the
path-walks are found to be strongly diverse, we can conclude that the
the progression of the agent is characterized by great freedom of
choice and variety of transient dynamics.  Consequently, because the
path-walks can not repeat nodes, we also have that the self-avoiding
walks in this case will also involve several diverse nodes.
Therefore, nodes with high diversity constitute natural candidates as
distributing sources (e.g. for information or mass).  It is also
important to observe that the dynamics diversity of a node provides
information which is complementary to other measurements of network
nodes.  For instance, though diversity tends to be correlated with the
node degree at the initial steps of the self-avoiding walks, such a
correlation can be quickly lost as a consequence of the structural
diversity at the progressive surroundings of the initial node.  The
walk diversity is also distinct from the betweeness centrality
(e.g.~\cite{Newman:2003,Costa_surv:2007}) in the sense that chained
nodes with high betweeness centrality will lead to low walk diversity.
Therefore, diversity can be best thought as a novel measurement which
can complement previous approaches in the characterization of the
structural properties of complex networks.

Many are the interesting applications of such diversity studies to
real-world problems.  For instance, in case the random walks are used
to model the acquisition of knowledge or cultural values by the agent
(in this case each node represent a knowledge or cultural fact,
e.g.~\cite{Costa_know:2006}), the diversity measurements can provide
sound basis for discussing how diverse the development of agents
starting from similar backgrounds but subsequently exposed to
different information will be.  Another particularly interesting
application concerns the objective quantification of the diversity of
life and species along phylogenetics, as well as geographical
exploration.  In addition to its potential for the objective
characterization of dynamics performed in complex networks, the
quantification of the diversity of random walks and path-walks can
also provide valuable indications about the structure of the
respective networks.  For instance, in case all path-walks are
identical, we have a chain of nodes extending from the starting node.
Contrariwise, a high diversity implies the presence of redundancies in
the network.

Because the diversity entropies at subsequent steps tend to be
correlated, the statistical method known as \emph{principal component
analysis} (PCA)~\cite{Costa_book:2001} is systematically applied in
this work in order to decorrelate those measurements.  The PCA method
provides an optimal stochastic linear transformation in the sense of
concentrating the variation of the data along the first new random
variables.  In other words, the PCA transforms the original
measurements into new features which are completely uncorrelated one
another.  Because of the linear nature of PCA, the new obtained
measurements correspond to linear combinations of the original
features, weighted so as to optimize the concentration of variance
along the first new variables.

The manuscript starts by presenting the basic concepts, adopted
network models, as well as the definition of the diversity entropy and
some of its properties.  The results are presented with respect to the
analysis of whole network categories, individual networks, and
individual nodes.

\section{Basic Concepts}~\label{sec:basic}

This section describes the basic concepts and methods used in this
article, including network representation and characterization, the 6
adopted complex network models, the definition and estimation of the
diversity entropy signature, as well as the stochastic projections
methods applied in order to decorrelated the signatures and achieve
dimensionality reduction.

\subsection{Complex Networks Representation and Characterization}

A unweighted and undirected complex network, formed by $N$ nodes and
$E$ edges, can be fully represented in terms of its
\emph{adjacency matrix} $K$, which is symmetric and has dimension $N 
\times N$.   Each existing edge $(i,j)$ implies $K(i,j)=K(j,i)=1$, with
$K(i,j)=K(j,i)=0$ indicating absence of that edge.  Two edges are said
to be \emph{adjacent} whenever they share one of their extremities.  A
\emph{random walk} corresponds to any sequence of 
adjacent edges $(i_1,i_2); (i_2,i_3); \ldots (i_{p-1},i_p)$. A walk
which does not repeat any edge or node, henceforth called
\emph{self-avoiding random walk}, defines a \emph{path} in the network.  
The \emph{length} of a walk or path is equal to the number of its
constituent edges.  The shortest path between two nodes is defined as
one of the paths between those nodes which has the smallest length.

The \emph{immediate neighbors} of a node $i$ are those nodes which are
connected to $i$ through shortest paths of length 1. The \emph{degree}
of a node is equal to the number of edges emanating from that node.
The node degree averaged within a network is called its \emph{average
degree}.  \emph{Extremity nodes} are henceforth understood as those
with unit degree.  As such, extremity nodes tend to determine the
termination of many self-avoiding walks (no way back for the moving
agent from that type of nodes). The \emph{clustering coefficient} of a
node $i$ is the ratio between the number of undirected edges between
the immediate neighbors of $i$ and the maximum possible number of
undirected edges among those nodes.

\subsection{Complex Networks Models}

Six theoretical models of complex networks are considered in the
present work including four traditional models --- Erd\H{o}s-R\'enyi
(ER), Barab\'asi-Albert (BA), Watts-Strogatz (WS) and a geographical
model (GG) --- as well as two recently introduced knitted types of
complex networks~\cite{Costa_comp:2007} --- the path-transformed BA
model (PA) and path-regular networks (PN).  The ER, BA and WS networks
are grown in the traditional way (e.g.~\cite{Albert_Barab:2002,
Dorogov_Mendes:2002, Newman:2003, Boccaletti:2006, Costa_surv:2007}).
The GG networks in this work are obtained by distributing $N$ nodes
within a square with uniform probability and connecting all nodes
which are closer than a minimal distance $d$.  The PA and PN networks
are obtained as explained in~\cite{Costa_comp:2007}: the PA networks
(\emph{path-transformed BA} networks) are obtained by star-path
transforming all nodes in an original BA network and the PN
(\emph{path regular} networks) model is easily obtained by defining
paths involving all network nodes in random order and without
repetition.

All networks considered in this article have $N \approx 100$ and $m
\approx 3$ or $m \approx 5$ ($m$ is the number of spokes in the added
nodes in the BA model), with average degree $\left< k \right> \approx
2m$.  The approximations are a consequence of the statistical
variability of the models.  For the same reason, the number of nodes
$N$ can vary slightly for the GG networks.  Because the average
degrees considered in this work are relatively large (well above the
percolation critical value for ER), most of the nodes in each network
belong to the largest connected component, which has been considered
for all the analyses reported in this article.  The total of
rewirings used in the WS case was equal to $0.1E$.

\subsection{Diversity Entropy and its Estimation}

The \emph{diversity entropy} is the measurement used in this article
in order to quantify the diversity of the self-avoiding random walks
obtained for each node $i$ at each step $h$.  Let $p(i,j,h)$ be the
probability that a node $j$ be visited after $h$ time steps while
moving from the starting node $i$. Once a self-avoiding walk is
terminated (i.e. the moving agent can proceed no further), the moving
agent is understood to remain at the final node and contribute to the
probabilities and diversities for all remaining steps.  The diversity
entropy of node $i$ can now be defined as:

\begin{equation}
  E(i,h) = - \sum_{j=1}^{N} p(i,j,h) log(p(i,j,h))
\end{equation}

Given the starting node $i$, $E(i,h)$ can be understood as the
diversity entropy \emph{signature} for that node (see
Figure{fig:features}).  Each such signature can be transformed into a
single value, e.g by taking the arithmetic or geometric average of its
values considering all the steps $h$.

\begin{figure}
  \vspace{0.3cm}
  \begin{center}
  \includegraphics[width=0.9\linewidth]{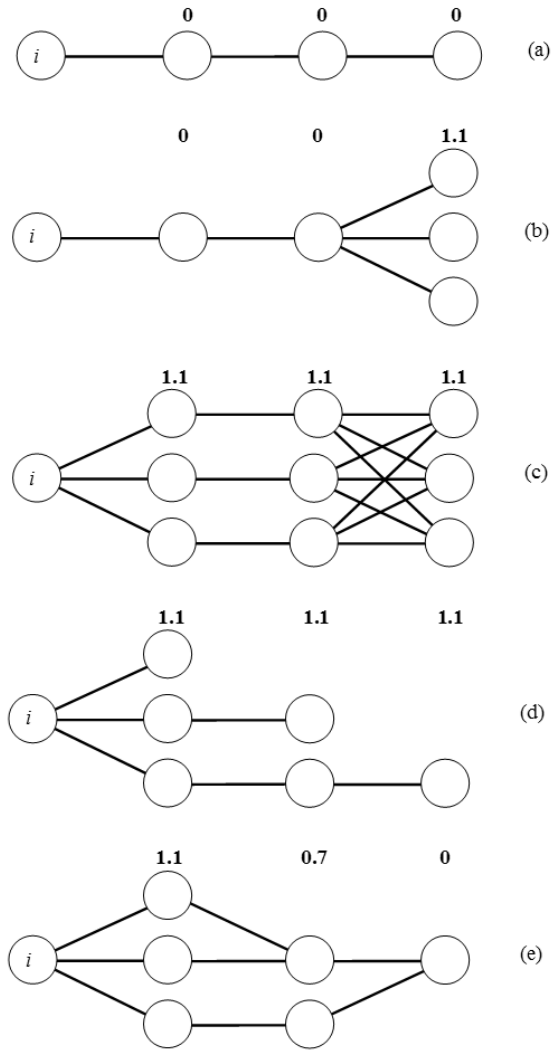}
  \caption{Illustrations of diversity entropy signatures
     (the diversity entropies are shown in bold):
     (a) as the paths from node $i$ are all equal for this case,
     the entropies are null for all values of $h$;
     (b) the divergence of edges at $h=3$
     implies the increase of the diversity entropy to 
     $1/log(3) \approx 1.1$ at that level;
     (c) the presence of all-to-all connections between
     the nodes at steps 2 and 3 has no effect in
     increasing the entropies at level $h=3$;
     (d) because the moving agent remains at each terminal node after
     reaching it, the entropy does not change at the successive steps
     for this case;
     (e) converging edges (at the second and third steps 
     in this particular example) can lead to
     decrease of the diversity entropy along $h$.}~\label{fig:ex_divers}
  \end{center}
\end{figure}

Figure~\ref{fig:ex_divers} illustrates several particularly relevant
situations regarding diversity entropy signatures.  Because of the
total absence of branches, the chain network in (a) yield a completely
null diversity signature.  This means total determinism in the sense
that all self-avoiding random walks starting from $i$ will be
identical. The presence of a branch at step 3 in the structure in (b)
implies the increase of the diversity entropy at this specific step.
In the network in (c), the branch occurs at the first step, implying
diversity entropy $E(i,1) = log(1/3) \approx 1.1$, which remains for
the two following steps (i.e. $h=2$ and 3).  Observe that the
additional all-to-all connections between the nodes in the second and
third steps have no effect in changing the respective diversity
entropy, as they do not affect the respective probabilities
$p(i,j,3$).  The situation depicted in (d) involves self-avoiding
random walks with different lengths, namely 1, 2 and 3.  Because the
moving agent is assumed to remain at its termination node, the
diversity entropies do not change along the 3 initial steps.  Though
this assumption implies eventual degeneracy such as obtaining the
same diversity entropy signatures for the structures in (c) and (d),
the distinction between such cases can be easily accomplished by
considering additional measurements such as the length of the walks.
Finally, the situation shown in (e) involves converging connections at
steps 1 and 2, which contribute to reducing the diversity of the
random walks.  Observe that the alternative assumption of removing the
moving agent after it has reached a termination node would imply
identical diversity entropy signatures for both structures in (d) and
(e).

Given a network with $N$ nodes, the maximum diversity entropy obtained
at any step is given when $p(i,j,h) = 1/N$, implying

\begin{equation}
  W = - 1/N \sum_{j=1}^{N} log(1/N) = log(N)
\end{equation}

\begin{figure}[htb]
  \vspace{0.3cm}
  \begin{center}
  \includegraphics[width=0.9\linewidth]{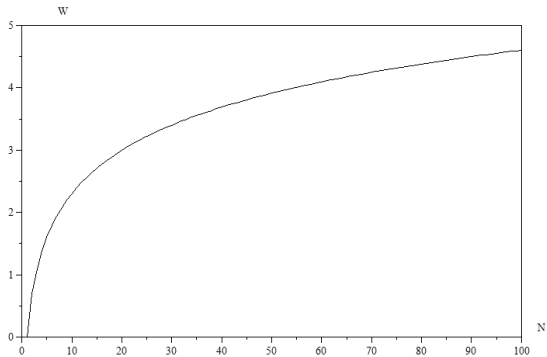}
  \caption{The maximum diversity entropy which can be obtained
        for networks with $N$ nodes.}~\label{fig:E_max}
  \end{center}
\end{figure}

Figure~\ref{fig:E_max} shows the maximum diversity entropies for
several values of $N$.  Therefore, as all networks considered in this
article involves $N \approx 100$, the diversity entropy is maximally
bound to $W = log(1/100) \approx 4.61$.

\begin{figure}[htb]
  \vspace{0.3cm}
  \begin{center}
  \includegraphics[width=0.9\linewidth]{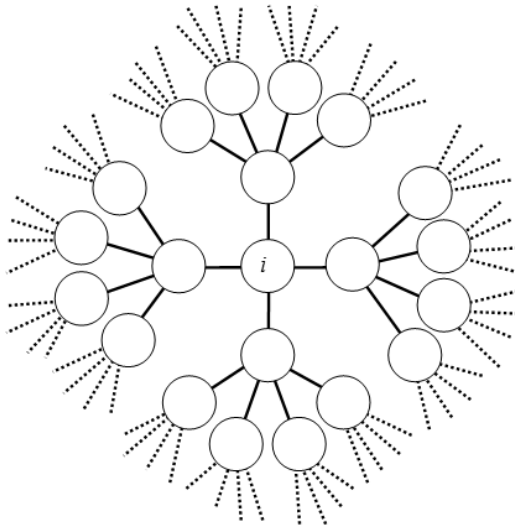}
  \caption{An example of a situation where the diversity entropy
     increases linearly with the steps $h$.}~\label{fig:special}
  \end{center}
\end{figure}

A particularly interesting situation occurs when each node at each
level $h$ leads exclusively to a constant number $\left< k \right>$ of
new nodes in the subsequent level $h+1$ (see
Figure~\ref{fig:special}). In this case, $p(i,j,h) = 1/(\left< k
\right>^{h})$, so that

\begin{equation}
  E(i,h) = - \sum_{j=1}^{\left< k \right>^h} log(\left< k \right>^{h})/\left< k \right>^{h} = h log(\left< k \right>)
\end{equation}

Therefore, the diversity entropy will tend to increase (or remain null
for $\left< k \right>=1$) with $h$ at constant rate $log(\left< k
\right>)$.  This situation involves an infinite and completely regular 
network (i.e. each node has the same degree $\left< k+1 \right>$).  As
complex networks are often analyzed with respect to regular or nearly
regular counterparts (e.g. ER model), it is useful to consider the
above configuration as a reference.  For instance, the situation in
which the diversity entropy tends to increase almost linearly with $h$
along an interval can be understood as an indication that the network
is mostly regular along that interval.  However, it should be born in
mind that linear increase of the diversity entropy can also be caused
by other structural organizations in complex networks (i.e. constant
increase of entropy does not necessarily implies network degree
regularity, but the latter necessarily implies linear entropy
increase).

As the diversity entropy has been defined for each node $i$ at each
step $h$, it provides an individual signature associated to each node
(see Figure~\ref{fig:features}), which can be valuable while
investigating dynamics emanating from that node.  However, it is often
interesting to get an overall idea of the diversity entropy dynamics
considering all the nodes in the networks.  This can be immediately
obtained in terms of the average and standard deviation of the
diversity entropy at each step $h$, i.e.:

\begin{eqnarray}
  \left< E(h) \right> = \frac{1}{N} \sum_{i=1}^{N} E(i,h) \\
  Var\{E(h)\} = \frac{1}{N} \sum_{i=1}^{N} ( E(i,h) - \left< E(h)
  \right>)^2 \\ \sigma_{E(h)} = +\sqrt{Var\{E(h)\}}
\end{eqnarray}

The algorithm adopted for picking a random path is simple and, given
each starting node $i$, involves performing $M$ self-avoiding random
walks along the initial $S$ steps (in this work, $S=10$). Therefore,
for each node $i = 1, 2, \ldots, N$, an accumulator array $V$ of
dimension $S \times N$ is kept in order to store the number of visits
to each of the network nodes after starting from $i$.  At every step
$h = 1, 2, \ldots, S$ along each of these $M$ self-avoiding walks, the
moving agent is found at a node $j$ and the element $V(h,j)$ of the
accumulator vector is incremented by one.  After completing the $M$
self-avoiding walks starting from node $i$, the probability of visits
to nodes can be estimated as $p(i,j,h) = V(h,j) / M$.  Recall that
once the moving agent reaches a termination node, it remains there for
all remaining steps.  For $N=100$, the situation considered for all
networks in this article, the probabilities have been experimentally
found to have converged to less than $5\%$ stability for $M=200$,
which is henceforth adopted.

\subsection{Optimal Dimensionality Reduction by Stochastic 
Transformations}~\label{sec:stat}

In several situations, especially for nearly uniform network
(i.e. nodes having most nodes with similar properties, such as
degree), the diversity neighboring entropies along the steps of the
signatures obtained for each node $i$ will tend to be strongly
correlated.  Indeed, recall that each node $i$ in the network will be
mapped into a $10-$dimensional feature vector (the diversity entropy
signature), which is a relatively high dimensional space, impossible
to be visualized.  It is possible, and useful, to reduce the
dimensionality of such measurement spaces by using optimal stochastic
linear transformations such as \emph{principal component analysis} and
\emph{canonical projections} (e.g.~\cite{Costa_surv:2007,Costa_book:2001}).  

The former of these approaches allows the measurement space to be
optimally projected into $m \leq S$ dimensions, $m \geq 1$, while
maximizing the variation of the observations along the first
transformed, new variables
(e.g.~\cite{Costa_surv:2007,Costa_book:2001}). The transformed
variables, which are linear combinations of the original measurements,
are guaranteed to be completely \emph{uncorrelated}.  The latter
transformation (i.e. canonical projections) allow the measurement
space to be optimally projected into a smaller dimensional space while
maximizing the separation of the categories of observations, in the
sense of maximizing the interclass variation and minimizing the
intraclass dispersion (e.g.~\cite{Costa_surv:2007,
McLachlan:1998}).  In both cases, the resulting transformed
variables are \emph{linear combinations} of the original measurements.

Though still rarely applied in complex network research, such optimal
stochastic transformations can be a real help in organizing and
simplifying the analysis and classification of complex networks
(e.g.~\cite{Costa_surv:2007}).  In this work, the principal component
approach is used in order to decorrelate the diversity entropy
signatures obtained for each of the nodes in a given complex network,
while the canonical projections method is applied in order to obtain
visualizations of the distribution of several realizations of 6
different types of complex network models.  Additional information
about the canonical projections analysis, which is mathematically more
sophisticated than the principal component approach, can be found
in~\cite{McLachlan:1998, Costa_surv:2007}.  The principal component
analysis, used to decorrelated the diversity entropies in this work,
is described as follows.

Let $\vec{f}$ be the feature vector containing the $S$ measurements
obtained for each observation $i = 1, 2, \ldots, N$.  In the present
work, each node $i$ (an observation) is mapped into a diversity
entropy signature (the feature vector) of dimension $S
\times 1$.  The elements $C(i,j)$, $i,j = 1, 2, \ldots, S$ of the
\emph{covariance matrix} of such a dataset can be estimated as

\begin{equation}
  C(i,j) = \frac{1}{N-1} \sum_{p=1}^{N} (v(i) - \mu_{i})(v(j) - \mu_{j})
\end{equation}

where $\mu_{a}$ is the average of $v(a)$, $a = 1, 2, \ldots, S$.
Observe that $C(i,j) = Var(i)$ whenever $i=j$.  Also, we have that the
covariance matrix $C$ is necessarily symmetric.

Let $\gamma_i$, $i = 1, 2, \ldots, S$ be the eigenvalues of the
covariance matrix, ordered so that $\gamma_1 \geq \gamma_2 \geq
\ldots \geq \gamma_S$, and let $\vec{q_i}$ be the respectively 
associated eigenvectors.  The principal component analysis can be
obtained by performing the following stochastic linear transformation

\begin{equation}
  \vec{g} =  \left [ \begin{array}{ccc}
              \longleftarrow  &  \vec{q_1}  &  \longrightarrow  \\
              \longleftarrow  &  \vec{q_2}  &  \longrightarrow  \\
              \ldots          &  \ldots     &  \ldots  \\
              \longleftarrow  &  \vec{q_m}  &  \longrightarrow  \\
             \end{array}   \right]   \vec{f}
\end{equation}

where $m \leq S$, $m \geq 1$, $\vec{f}$ and $\vec{g}$ have respective
dimensions $S \times 1$ and $m \times 1$.  Therefore, the new
measurements (transformed variables) belong to a space of reduced
dimensionality $m \leq S$.  The new measurements associated to the
largest eigenvalues are called the \emph{main} variables or
components.  Observe that each of the new measurements is a linear
combination of the original measurements, while the eigenvalues
$\gamma_i$, $i = 1, \ldots, m$, correspond to the variances of the new
measurements in $\vec{g}$.  So, it is reasonable to include in the
transformation matrix only the eigenvectors associated to eigenvalues
which are particularly large, in order to encompass the greatest part
of the original variation of the observations.  Also, observe that the
relative weight of each of the original measurements used in the
linear combinations defining the new variables can provide an
indication about the importance of the respective original
measurements.  Because the feature vectors considered in this work
have all the same nature and potential dynamic range (recall that all
the elements of the diversity entropy signature are entropies vary
between 0 and log(N)), there is no need for preliminary
standardization of the original measurements
(e.g.~\cite{Costa_surv:2007}).

\section{Results and Discussion}

The diversity entropy methodology has been applied at the level of
network categories and individual nodes.  In the former case, the
overall diversity was characterized with respect to the average and
standard deviation of the diversity entropies obtained for each
realization of the networks.  The latter investigation targets the
estimation of the diversity entropy signature at the individual node
level, which allows the partitioning of each network into subgraphs of
similar diversity.  These two types of investigations, at the network
and individual node levels, are described in the respective following
sections.

\subsection{Network Level}

\begin{figure*}[htb]
  \vspace{0.3cm} 
  \begin{center}
  \includegraphics[width=0.9\linewidth]{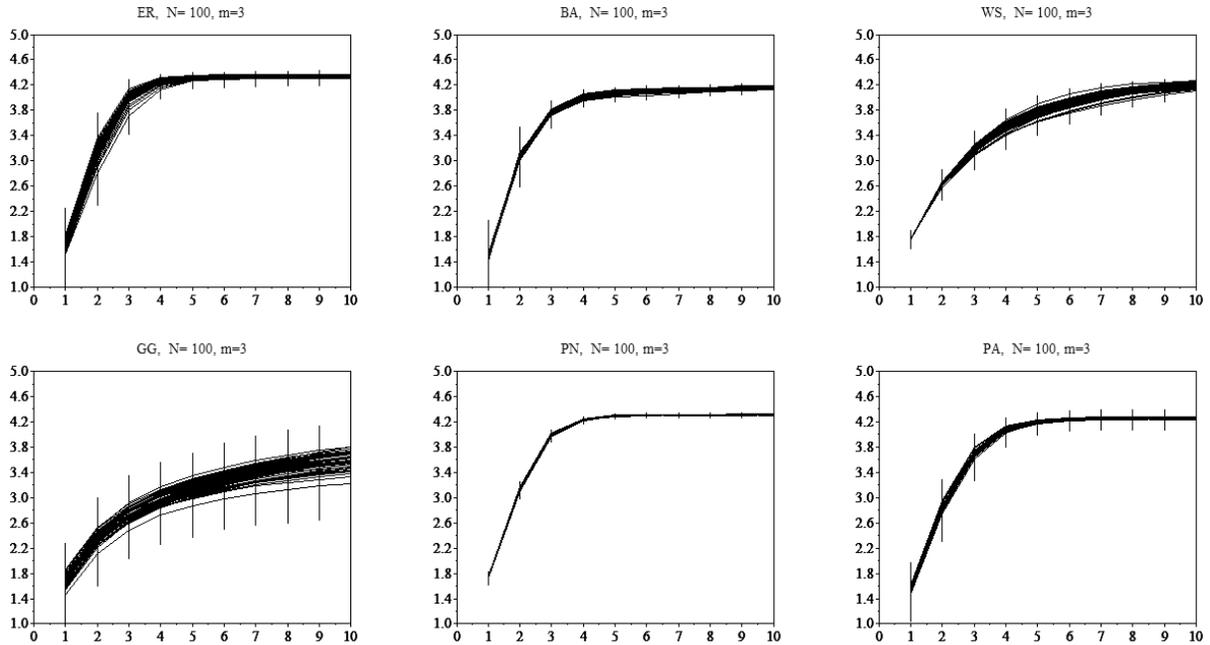} 
  \caption{The average $\pm$ standard deviations of 
      the diversity entropies obtained for each of the networks
      considered for each of the six complex networks models 
      assuming $N=100$ and $m=3$ (i.e.  
      $\left< k \right> = 6$).}~\label{fig:signts_by_network_3} 
  \end{center}
\end{figure*}

We start our diversity investigation by looking at the averages and
standard deviations of the diversity entropy signatures obtained for
the realizations of each of the 6 considered network models.  More
specifically, a total of 50 realizations were performed for each of
the six complex network models considering $N=100$ and two average
degrees: (a) $\left< k \right> = 6$ (i.e. $m=3$) and (b) $\left< k
\right> = 10$ (i.e. $m=5$).  For each of such realizations, 200 random 
path-walks were performed starting from each of the nodes, and the
respective entropies $E(i,h)$ were estimated for $h = 1, 2, \ldots,
10$.  The average $\left< E(h) \right>$ and standard deviations
$\sigma_{E(h)}$ of these diversity entropy values were obtained for
each of the 50 network realizations for each of the 6 considered
models and are shown in Figures~\ref{fig:signts_by_network_3}
and~\ref{fig:signts_by_network_3} respectively to $m=3$ and $m=5$.

\begin{figure*}[htb]
  \vspace{0.3cm} 
  \begin{center}
  \includegraphics[width=0.9\linewidth]{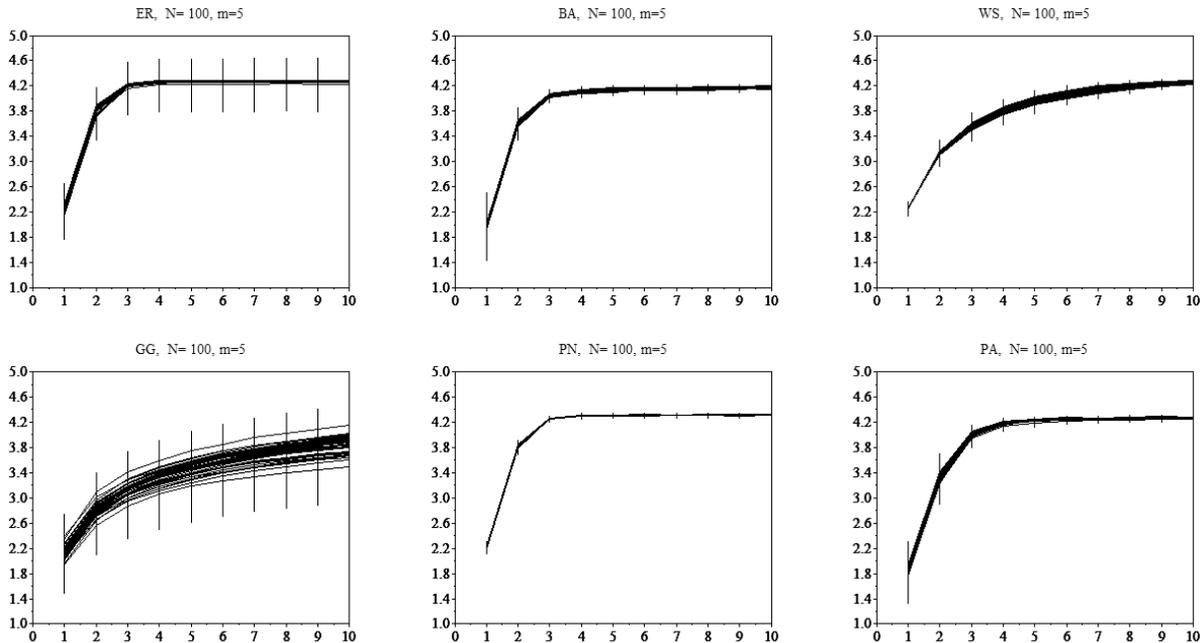} 
  \caption{The average $\pm$ standard deviations of 
      the diversity entropies obtained for each of the networks
      considered for each of the six complex networks models 
      assuming $N=100$ and $m=5$ (i.e.  
      $\left< k \right> = 10$).}~\label{fig:signts_by_network_5} 
  \end{center}
\end{figure*}

A series of interesting results can be inferred from the curves in
those figures.  First, observe that markedly distinctive curves and
standard deviations were obtained for the networks belonging to each
of the considered models.  Two general behaviors can be distinguished
in both figures: the steeper increase of the diversity entropy with
$h$ obtained for the ER, BA, PN and PA modes as opposed to the more
gradual increase verified for the WS and GG models.  These two types
of transient dynamics can be observed for both $m=3$ and $m=5$.  In
the case of the transient evolutions observed for the ER, BA, PN and
PA models, the diversity entropy tended to reach stabilization near a
higher plateau (with diversity entropy approximately equal to 4.2)
after the three or four initial steps.  This suggests that the
self-avoiding paths in these networks tend to reach most nodes after
just a few steps.  The more gradual increase of entropy observed for
the WS and GG models indicates that the moving agent takes
substantially more time to cover a smaller portion of the nodes during
the transient dynamics.  This is a consequence of the fact that,
though nearly regular (i.e. similar degrees for all nodes), these two
types of networks are characterized by having pairs of nodes which are
either connected through many short paths (adjacent nodes) or
virtually unconnected.  More informally, given two nodes $i$ and $j$
of a network, the \emph{adjacency} between them can be quantified in
terms of the number of short (i.e. up to a maximum length) paths
interconnecting those nodes; the higher this number, the more adjacent
the pair of nodes is.

Another interesting result regards the maximum entropy values, reached
for large values of $h$.  Except for the GG model, all other types of
networks tended to entropies around 4.2 for both $m=3$ and $m=5$.
Interestingly, quite similar limiting entropy values have been
obtained for BA and PN networks irrespectively of the average degree.
By comparing the respective curves in the two figures, it becomes
clear that the higher average degree (i.e. $m=5$, implying $\left< k
\right> = 10$) tended to reduce the standard deviations along the plateaux 
of the BA, WS, PN and PA cases.  Similar standard deviations can be
observed for the GG case, while higher deviations now characterize the
plateaux for the ER models. Though at first surprising, the decrease
of the standard deviations observed for most models is ultimately a
consequence of the fact that increasing the average degree of finite
networks tends to make them more regular, implying in more
self-avoiding paths covering the same set of nodes.  Observe that at
the extreme situation in which the network is fully connected, all
path-walks will involve all nodes, implying null variance of the
diversity entropy.  The higher average degree also tended to change
the shapes of the curves by implying a steeper increase (especially
for the second step) along the initial step, which is also a
consequence of the above observed regularizing effect.

Another interesting result which is evident from
Figures~\ref{fig:signts_by_network_3}
and~\ref{fig:signts_by_network_3} are the markedly distinct standard
deviations obtained for each model.  Confirming previous
investigations~\cite{Costa_comp:2007,Costa_longest:2007}, the PN model
presented the more regular features, with almost null standard
deviations of the diversity entropies for either $m=3$ or $m=5$.
Allied with the fast increase of diversity entropy exhibited by this
model, the extremely low variance of diversity entropies makes of the
PN a choice model for achieving high and uniform diversity signatures.
To a great extent, such properties favoring diversity are a
consequence of the fact that, unlike the WS and GG models, any pair of
nodes in the PN structures tend not to be adjacent in the sense of
being interconnected by many short paths.  Recall that the ER, WS, GG
and PN are all network models characterized by high degree regularity,
so that what makes them so different regarding diversity is ultimately
the adjacency between pairs of nodes, which is optimally broken in the
PN model.

\begin{figure*}[htb]
  \vspace{0.3cm}
  \begin{center}
  \includegraphics[width=0.45\linewidth]{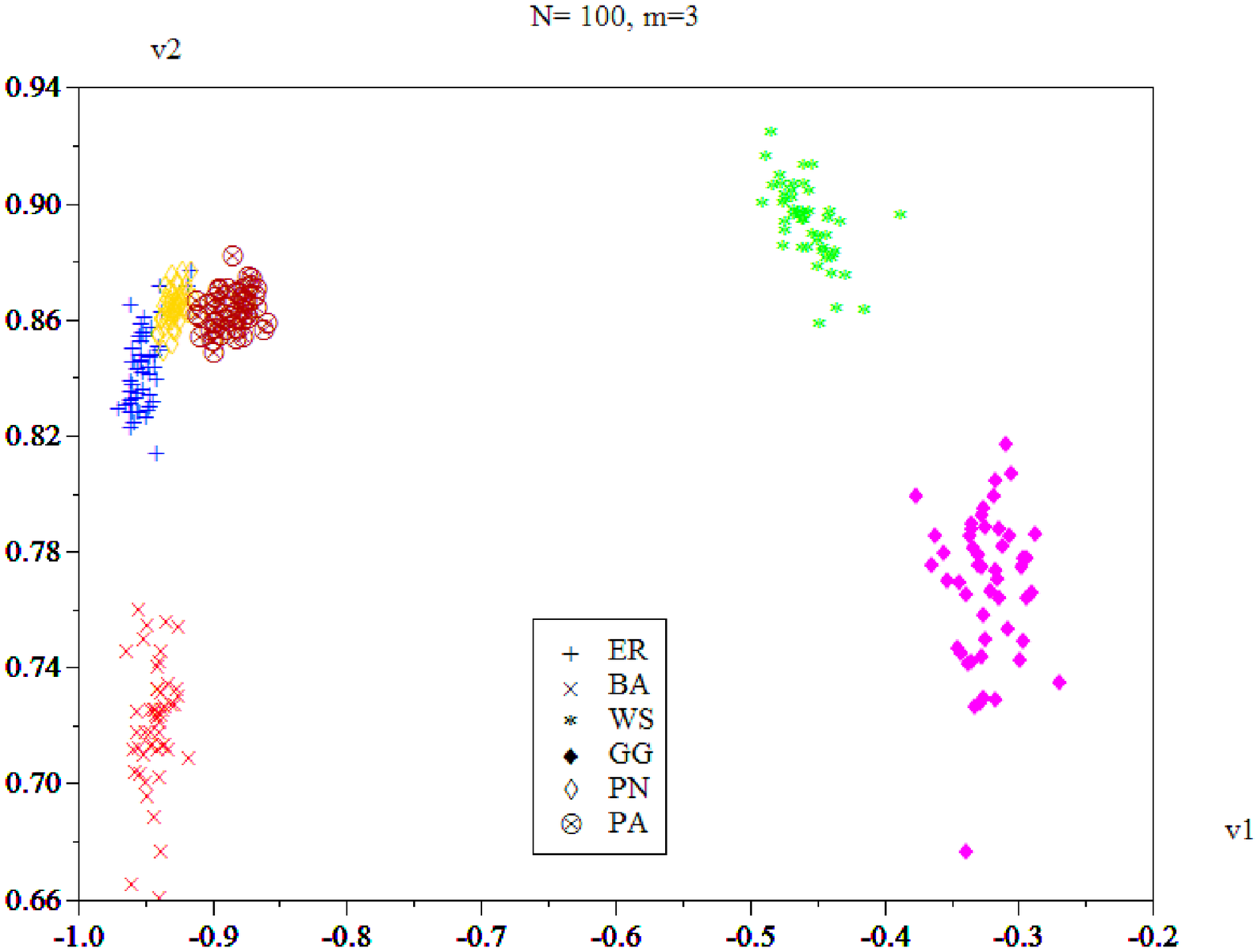} \hspace{1cm}
  \includegraphics[width=0.45\linewidth]{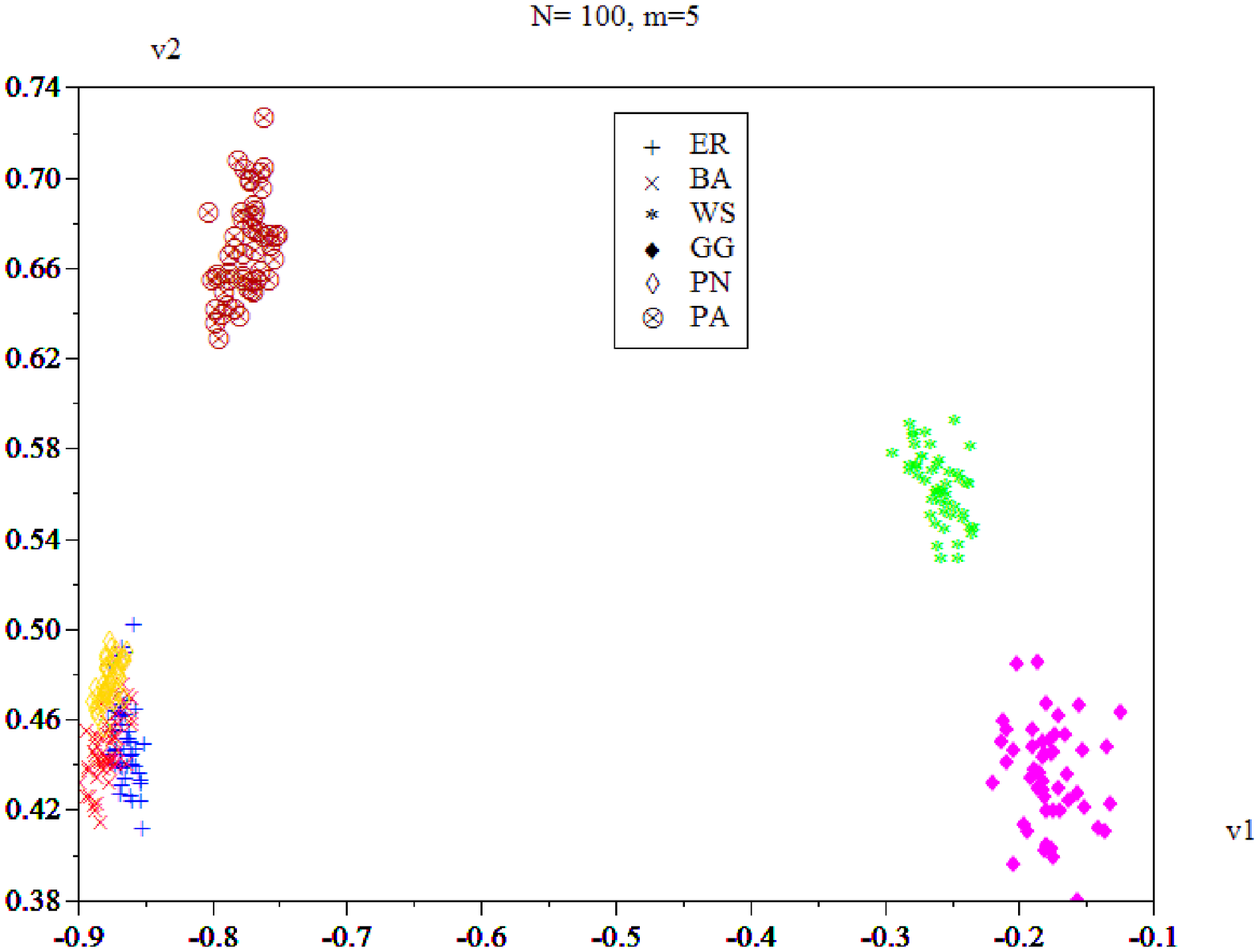} \\
  (a) \hspace{9cm} (b)
  \caption{The clusters of networks, for $m=3$ (a) and $m=5$ (b), 
      after being canonically projected from 10 to 2 
      dimensions so as to maximize the separation between
      the six categories of networks.}~\label{fig:canonical}
  \end{center}
\end{figure*}

In order to complete our analysis of the diversity entropy signatures
for networks belonging to the 6 distinct considered models, we now
apply the canonical projection method (see Section~\ref{sec:stat}).
We use this method to project the original $10-$dimensional entropies
space into a $2-$dimensional space so as to maximize the separation
between the clusters of networks belonging to each category.
Figure~\ref{fig:canonical} shows the cluster distributions obtained
for $m=3$ (a) and $m=5$. 

It is clear from Figure~\ref{fig:canonical}, where $v1$ and $v2$
correspond to the two principal canonical variables, that the 6
categories of networks yielded two supergroups: one formed by $\{GG,
WS\}$ and the other by $\{ER, BA, PN, PA\}$ (observe the different
ranges of values for the two axes).  This is in complete agreement
with the two main types of diversity dynamics identified for those
networks (i.e. steeper and more gradual increase of the entropies).
In addition to confirming that previous result, the canonical
projections showed that the ER and PN have marked similarity between
their diversities (i.e. the clusters for these two models were mapped
nearby in the projected space).  The smallest dispersion of the
diversities obtained for the PN model are clearly reflect in the dense
cluster obtained for that category of networks.  Interestingly, the ER
and PA clusters tended to change positions considerably for $m=3$ and
$m=5$.  

Additional results can also be obtained by considering the weights of
the original measurements in the linear combinations defining the two
canonical variables $v1$ and $v2$, shown in Table~\ref{tab:can}.  We
concentrate attention on the absolute values of the weights in
Table~\ref{tab:can}.  In the case $m=3$, we have that the two first
canonical variables $v1$ and $v2$ are mostly affected by the diversity
entropies for $h=1, 2, 4, 7, 8, 9$ and $10$, which are the main
measurements responsible for the optimal separation between the 6
models in the case $m=3$.  The most important measurements in the
composition of the two canonical variables for $m=5$ are the
diversities obtained for $h = 1, 2, 3, 4$, all of which with weights
larger than $0.5$.  These measurements, which correspond to the
initial steps of the walks, greatly contributed to the separation of
the 6 network categories.  The dominant contribution of the initial
entropies is reasonable because most the diversity signatures tend to
become stable and similar after 3 steps in the case of $m=5$ (recall
that the diversity entropies rise more abruptly for $m=5$ than for
$m=3$).  Such results obtained for $m=5$ indicate that, if the main
purpose of the analysis is to separate the 6 network types, it is
mostly enough to consider the 4 initial entropies in each signature.

\begin{table}[htb]
\centering
\begin{tabular}{|c|c|c|c|}  \hline  
   \multicolumn{2}{|c|}{$m=3$}   &  \multicolumn{2}{c|}{$m=5$}  \\ \hline
    $v1$   &     $v2$   &    $v1$    &    $v2$  \\ \hline
   0.45    &    0.37    &    0.51    &  0.12    \\ \hline     
   -0.03   &    -0.36   &    -0.29   &  -0.58   \\ \hline  
   -0.21   &    -0.01   &    -0.63   &  -0.14   \\ \hline  
   -0.52   &    0.18    &    -0.16   &  0.57    \\ \hline  
   -0.08   &    0.19    &    0.07    &  0.23    \\ \hline  
   -0.09   &    -0.36   &    0.31    &  0.37    \\ \hline   
   0.40    &    -0.05   &    -0.05   &  0.02    \\ \hline  
   0.32    &    0.39    &    0.22    &  -0.16   \\ \hline  
   0.18    &    0.42    &    0.19    &  -0.30   \\ \hline  
   -0.42   &    -0.45   &    -0.19   &  -0.12   \\ \hline  
\end{tabular}
\caption{The weights of the original measurements assigned by
          the canonical projections method in order to best
          separate the 6 categories of complex networks in
          the $2-$dimensional projections for $m=3$ and $m=5$.}\label{tab:can}
\end{table}

\subsection{Individual Node Level}

\begin{figure*}[htb]
  \vspace{0.3cm} 
  \begin{center}
  \includegraphics[width=0.9\linewidth]{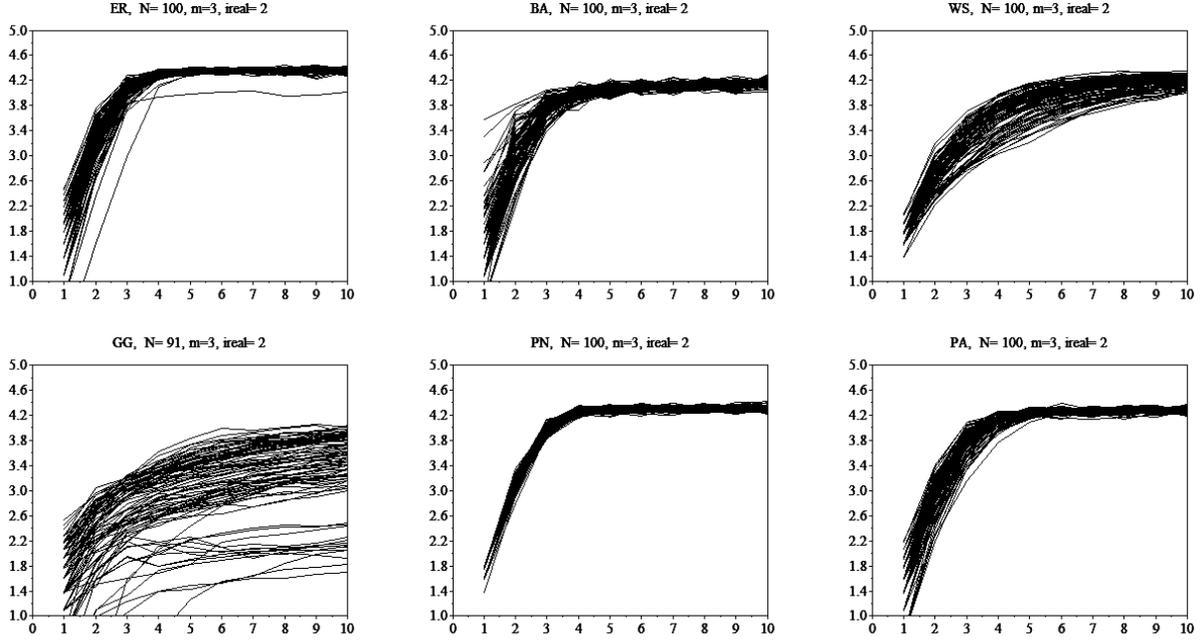} 
  \caption{The average $\pm$ standard deviations of 
      the diversity entropies obtained for each node in a sample
      of each of the six complex networks models 
      assuming $N=100$ and $m=3$ (i.e.  
      $\left< k \right> = 6$).}~\label{fig:signts_by_node_3} 
  \end{center}
\end{figure*}

Having investigated how the diversity entropies behave in each of the
6 considered network categories, we now turn our attention to the
diversity entropy signatures obtained at the level of individual
nodes.  In order to do so, we selected a network of each type and
obtained the respective signatures shown in
Figure~\ref{fig:signts_by_node_3} and~\ref{fig:signts_by_node_3}, with
respect to $m=3$ and $m=5$.  Similar signtures were obtained for other
realizations of each of the 6 types of networks.  The geographical
network considered in this analysis is shown in
Figure~\ref{fig:graph}.

\begin{figure*}[htb]
  \vspace{0.3cm} 
  \begin{center}
  \includegraphics[width=0.9\linewidth]{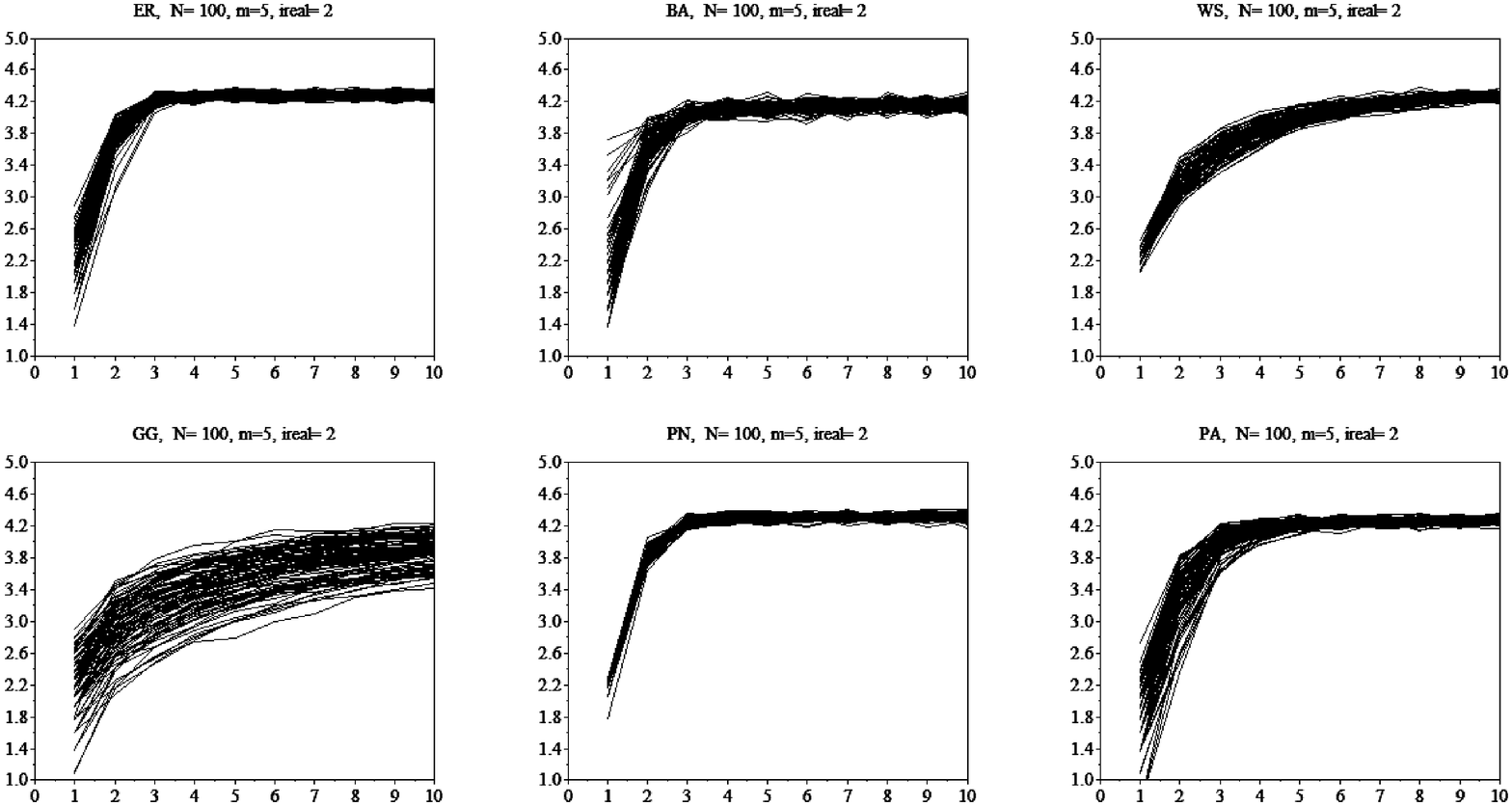} 
  \caption{The average $\pm$ standard deviations of 
      the diversity entropies obtained for each node in a sample
      of each of the six complex networks models 
      assuming $N=100$ and $m=5$ (i.e.  
      $\left< k \right> = 10$).}~\label{fig:signts_by_node_5} 
  \end{center}
\end{figure*}

Most of the results and explanations presented in the previous
analyses at the network level can be immediately extended to the
signatures in these curves.  First, the dispersion of the signatures
tend to be smaller for $m=5$ than for $m=3$.  Very similar signatures
were obtained for all nodes in the PN networks, which confirms the
distinctive regularity of this model.  The two types of transient
dynamics, namely steeper for the ER, BA, PN and PA networks and more
gradual for the WS and GG structures, were again observed.  The most
interesting additional information provided by the presentation of the
individual node signatures regards the relative dispersion obtained
for each case.  Observe that particularly distinct diversity
signatures were obtained for the GG structure.  This is mainly a
consequence of the higher structural modularity and overall adjacency
found in this type of network (see Figure~\ref{fig:graph}).

\begin{figure*}[htb]
  \vspace{0.3cm} 
  \begin{center}
  \includegraphics[width=0.9\linewidth]{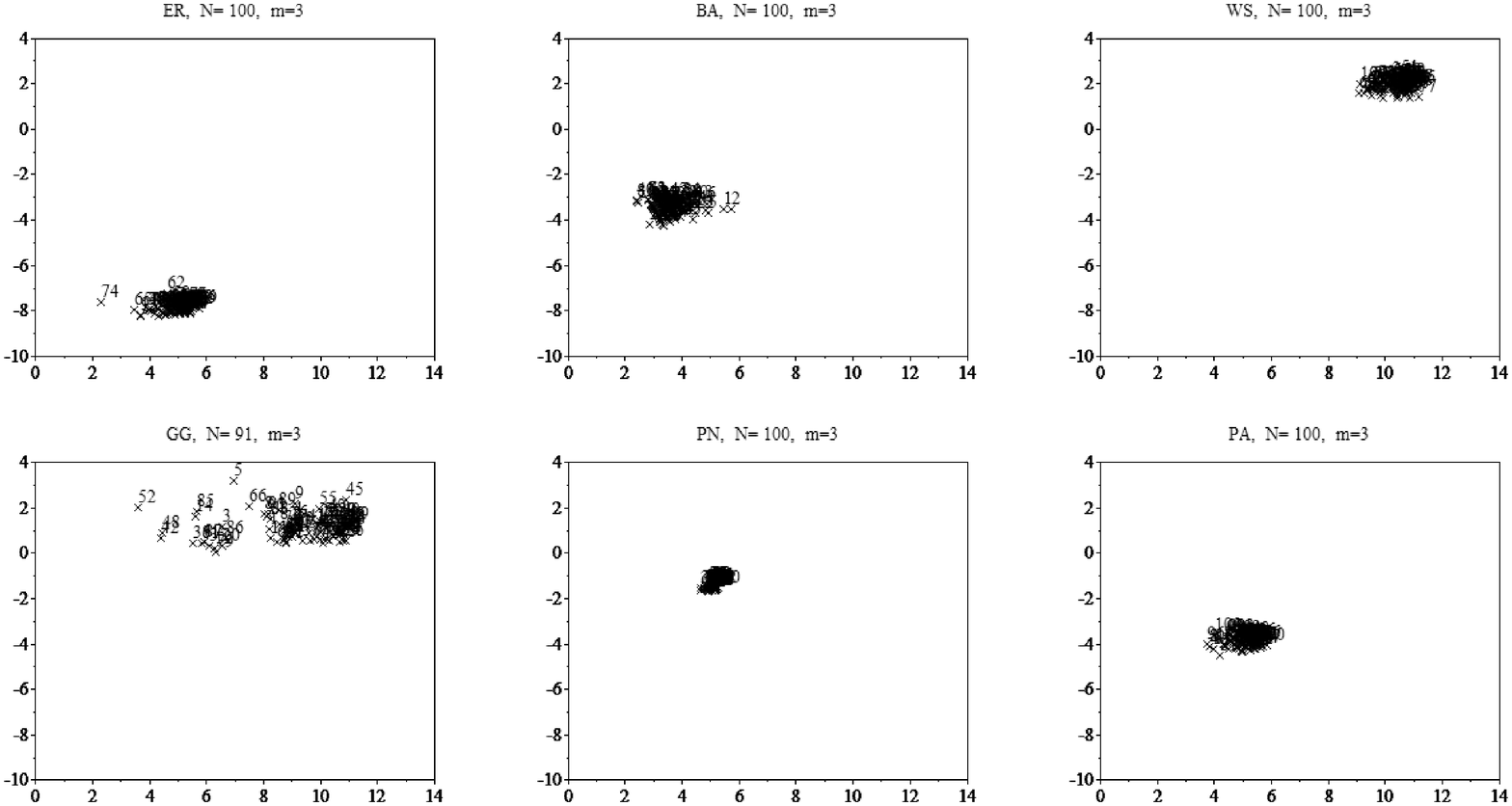} 
  \caption{The two-dimensional projections of the diversity entropy
      signatures obtained for each of the nodes in a sample of
      each of the six complex networks models assuming $N=100$ and 
      $m=3$ (i.e. $\left< k \right> = 6$).}~\label{fig:pca_3} 
  \end{center}
\end{figure*}

Additional insights about the measurement structure in each of the
networks can be obtained by applying principal component analysis to
each of the datasets in Figures~\ref{fig:signts_by_node_3}
and~\ref{fig:signts_by_node_3} in order to obtain $2-$dimensional
visualizations of the distribution of the respective diversity
signatures.  The projections obtained for $m=3$ is shown are
Figure~\ref{fig:pca_3} (the projections for $m=5$ are very similar and
are not shown in this article).  Recall that each point in these plots
corresponds to each of the nodes in the respective sample network.
Distinct clusters, all of which completely uncorrelated, were obtained
for each of the networks.  Note the presence of outliers for the cases
ER, BA and GG.  The distribution obtained for the GG structure
presented the largest dispersion of points.  Contrariwise, the
projection of the diversity entropy signatures for the PN network
yielded the most compact cluster, confirming once again the enhanced
regularity of this type of network.  The dispersions of the ER, BA, WS
and PA networks are similar.

Additional insights about the influence of the measurements (i.e. the
diversity entropies) in the definition of the clusters in
Figure~\ref{fig:pca_3} can be obtained by considering the respective
weights of the original measurements in the linear combinations
defining the two main principal variables $pca1$ and $pca2$, which
necessarily resulted completely uncorrelated.  Observe that the
variance of $pca1$ is much wider than that ofr $v2$ in all cases.
Such weights are given in Table~\ref{tab:pca}.  It is clear from the
values in this table that the initial 3 or 4 diversity entropies are
the most relevant measurements in all situations, except for the WS
and GG cases.  The special importance of the first entropies is a
consequence of the fact that the signatures tend to be more different
at such initial steps, getting nearly constant once they reach the
plateaux.  The distinct composition exhibited by the first principal
variable in the case of the WS and GG networks reflects that the
individual signatures remain distinct even after 3 or 4 steps.
Because of its largest variance, the first principal variables ire
particularly relevant as a single quantification of the individual
node diversities.  In the case of the WS and GG networks, this
variable corresponds very closely to the arithmetic mean (to a
multiplicative factor). This summarizing measurement is henceforth
called the \emph{overall diversity} of each node.

\begin{table*}[htb]
\centering
\begin{tabular}{|c|c|c|c|c|c|c|c|c|c|c|c|}  \hline  
   \multicolumn{2}{|c|}{ER}   &  \multicolumn{2}{c|}{BA}  &
   \multicolumn{2}{|c|}{WS}   &  \multicolumn{2}{c|}{GG}  &
   \multicolumn{2}{|c|}{PN}   &  \multicolumn{2}{c|}{PA}  \\ \hline
    $pca1$   &     $pca2$   &    $pca1$    &    $pca2$  &
    $pca1$   &     $pca2$   &    $pca1$    &    $pca2$  & 
    $pca1$   &     $pca2$   &    $pca1$    &    $pca2$    \\ \hline

   0.75    &    0.58    &    0.91    &  0.40  &   0.07    &   -0.50    &
   0.25    &   -0.73    &    0.52    &  0.44  &   0.73    &   0.65  \\ \hline  

   0.60    &   -0.52    &    0.38    & -0.89  &   0.33    &   -0.63    &
   0.30    &   -0.46    &    0.79    &  0.07  &   0.58    &   -0.41 \\ \hline  

   0.27    &   -0.38    &    0.13    & -0.19  &   0.47    &   -0.28    &
   0.31    &   -0.18    &    0.32    & -0.80  &   0.34    &   -0.60 \\ \hline  

   0.05    &   -0.23    &    0.03    & -0.03  &   0.47    &    0.08    &
   0.31    &    0.01    &    0.07    & -0.33  &   0.13    &   -0.22 \\ \hline     
   0.01    &   -0.20    &    0.02    & -0.06  &   0.42    &    0.23    &
   0.32    &    0.10    &    0.04    &  0.09  &   0.03    &   -0.06 \\ \hline     
   0.01    &   -0.17    &    0.02    &  0.01  &   0.33    &    0.27    &
   0.33    &    0.17    &    0.00    &  0.10  &   0.02    &   -0.03 \\ \hline     
   0.01    &   -0.14    &    0.01    & -0.07  &   0.26    &    0.26    &
   0.33    &    0.18    &   -0.03    &  0.12  &   0.01    &    0.00 \\ \hline     
   0.02    &   -0.18    &   -0.01    & -0.01  &   0.22    &    0.20    &
   0.33    &    0.21    &   -0.02    &  0.07  &   0.00    &   -0.02 \\ \hline  

   0.02    &   -0.21    &    0.03    & -0.02  &   0.17    &    0.13    &
   0.33    &    0.23    &    0.00    &  0.06  &   0.00    &    0.01 \\ \hline  

   0.01    &   -0.19    &    0.01    &  0.02  &   0.12    &    0.14    &
   0.33    &    0.24    &    0.04    &  0.06  &   0.01    &   -0.05 \\ \hline  \end{tabular}
\caption{The weights of the original measurements assigned by
          the principal component analysis method in order to 
          completely decorrelate the diversity entropy signatures
          for each of the networks representing each of the 6
          categories of networks.}\label{tab:pca}
\end{table*}

In order to conclude our investigation of the diversity entropy
signatures at the individual node level, we consider the GG network
chosen for the above examples (see Figure~\ref{fig:graph}) for a more
systematic investigation of the diversities.  The choice of this type
of network is justified because it is the only case among the
considered categories which incorporates the spatial positions of each
node and because this type of network tends to exhibit structured
modularity (i.e. spatial and topological communities).

\begin{figure*}[htb]
  \vspace{0.3cm} 
  \begin{center}
  \includegraphics[width=0.9\linewidth]{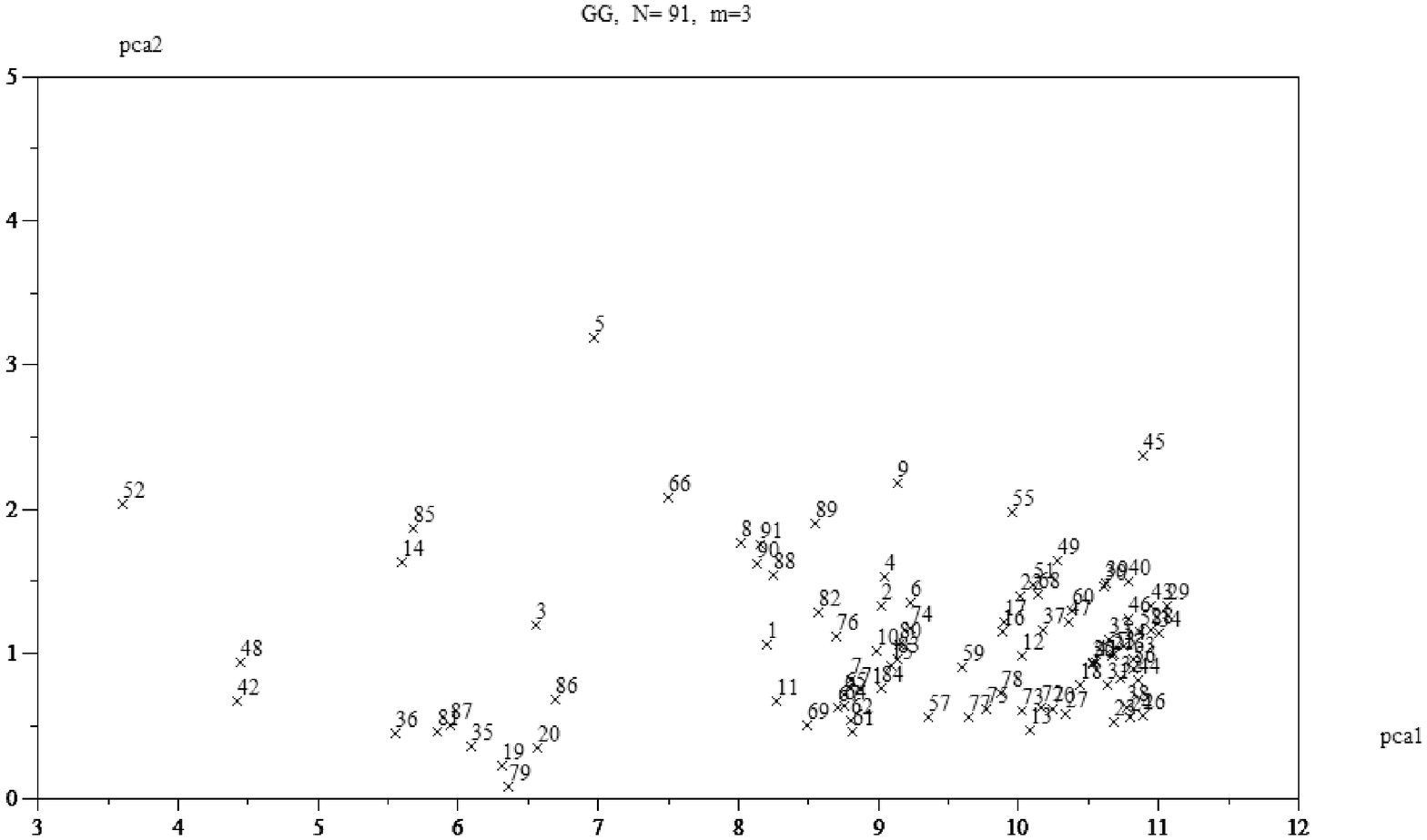}  \\
   (a)  \\  \vspace{0.5cm}
  \includegraphics[width=0.9\linewidth]{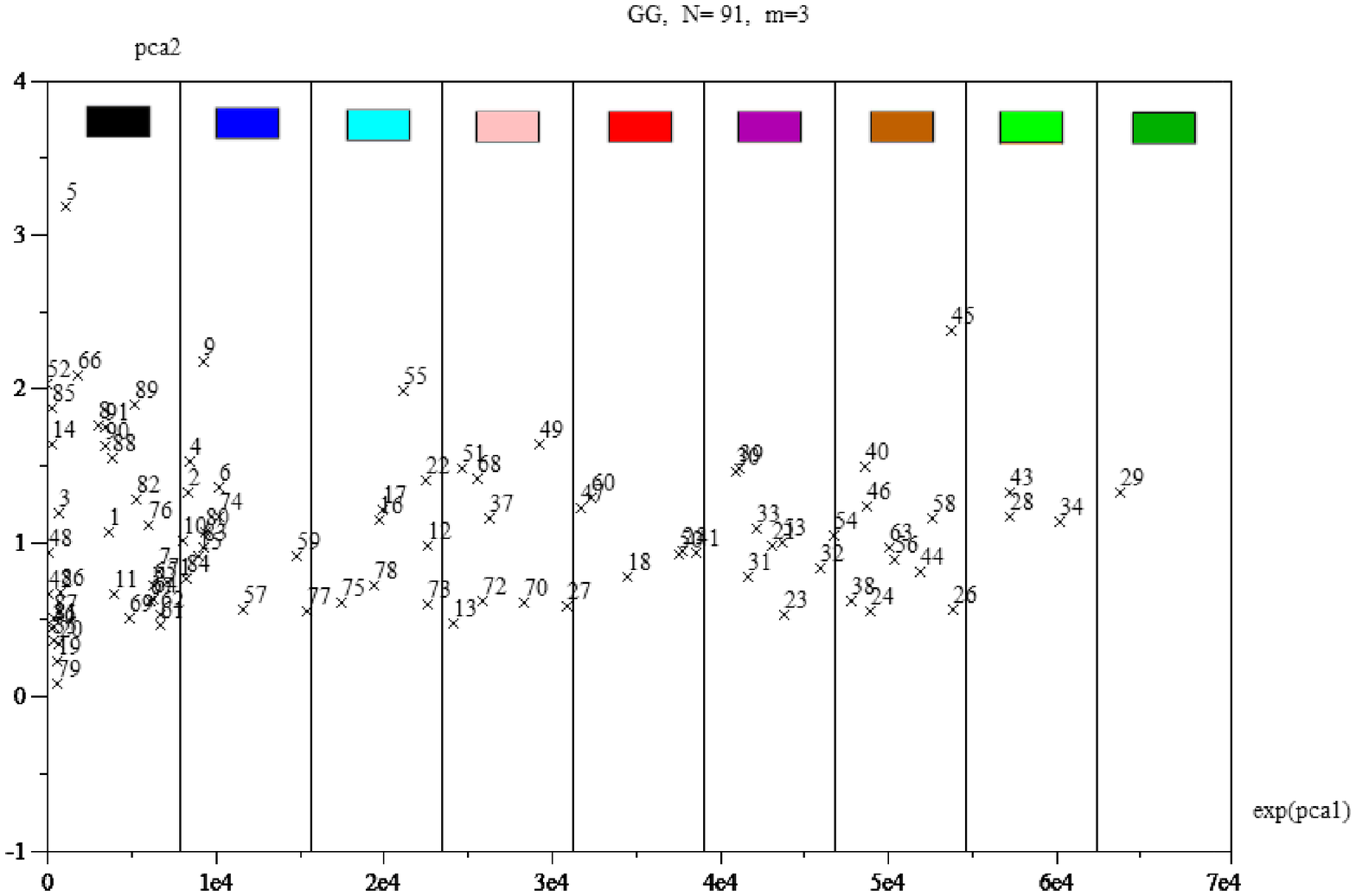} \\
   (b)
  \caption{The PCA projection obtained for the chosen GG network (a),
      and its transformation (b) obtained by taking the exponential 
      of the first PCA variable, i.e. $exp(pca1)$.  The colors
      identify each of the intervals of the $exp(pca1)$ range.
      }~\label{fig:pca_GG} 
  \end{center}
\end{figure*}

\begin{figure*}[htb]
  \vspace{0.3cm} 
  \begin{center}
  \includegraphics[width=0.7\linewidth]{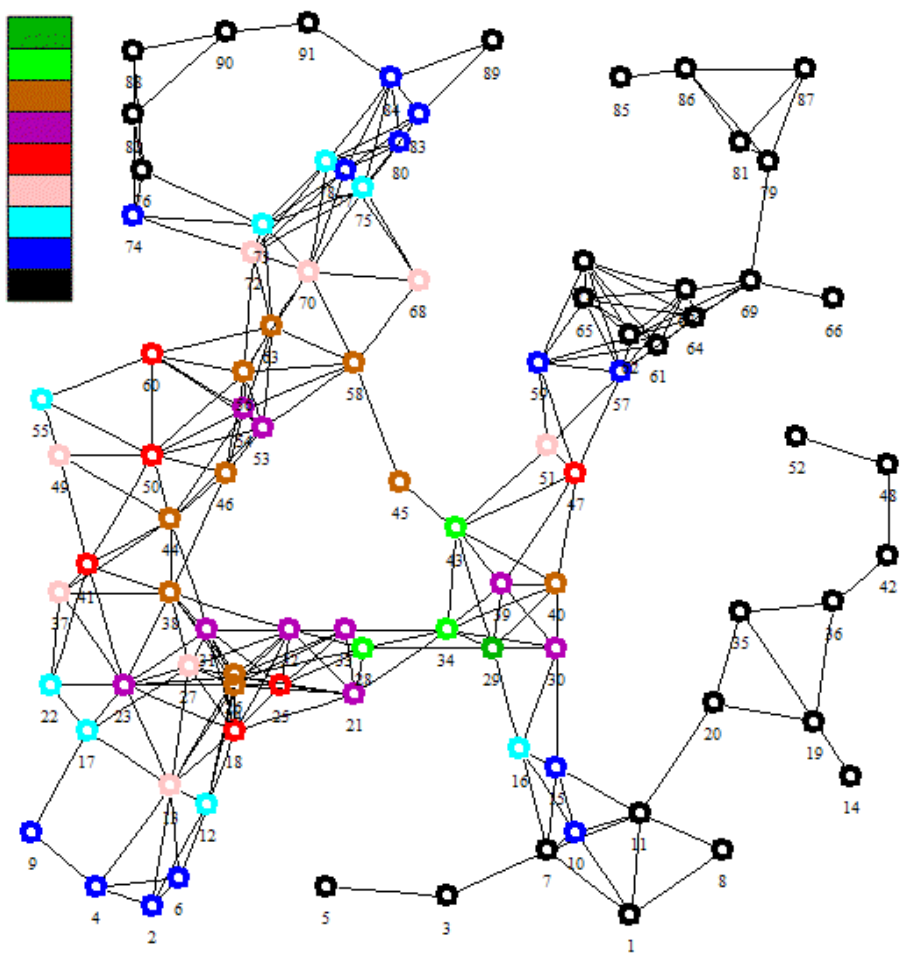} 
  \caption{The GG network with $n=98$ and $m=3$ and its nodes
       colored according to the intervals defined for the
       exponential of the first pca variable in 
       Figure~\ref{fig:pca_GG}.}~\label{fig:graph} 
  \end{center}
\end{figure*}

Because the first principal variable has been verified to correspond
very closely to the arithmetic average of the diversity entropies for
all network types, we adopt this value in order to summarize the
diversity of each node in the chosen network.
Figure~\ref{fig:pca_GG}(a) shows an enlarged version of the PCA
projection of the diversity entropies obtained for this geographical
network.  Because of the right-skewed distribution of the density of
the points in this projection, we consider a new projection obtained
by taken the exponential of the first PCA variable, henceforth
represented as $exp(pca1)$.  This new projected distribution is shown
in Figure~\ref{fig:pca_GG}(b).  A more uniform distribution of points
is now obtained.  We now subsume the new variable $exp(pca1)$ into 9
intervals identified by the colors in Figure~\ref{fig:pca_GG}(b).
Observe that, because $pca1$ is very close to the arithmetic average
of the diversity entropies for the various values of $h$, the
diversity of the nodes increase from left to right in both
Figures~\ref{fig:pca_GG}(a) and (b).  Figure~\ref{fig:graph} shows the
original GG structure with its nodes colored according to the
overall diversity intervals in Figure~\ref{fig:pca_GG}(b).  

\begin{figure*}[htb]
  \vspace{0.3cm} 
  \begin{center}
  \includegraphics[width=0.7\linewidth]{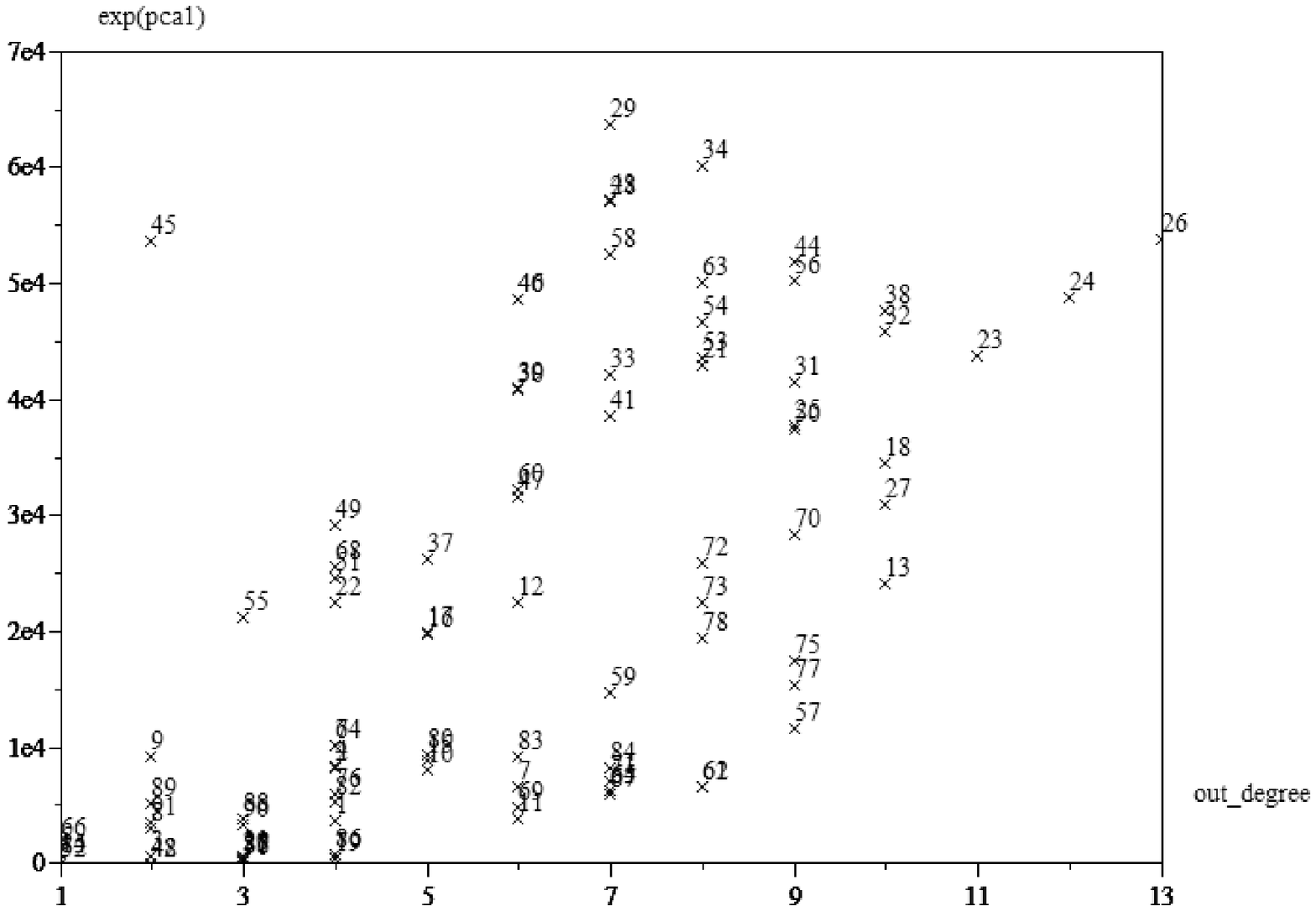} 
  \caption{The scatterplot obtained by considering the overall
        diversities of the nodes in the GG structure and ther
        respective degrees.  Only a weak correlation can be
        identified between these two measurements.}~\label{fig:corrs} 
  \end{center}
\end{figure*}

A series of interesting results can be identified.  First, observe
that the nodes belonging to more external structures tend to present
the smallest overall diversities (in black).  All extremity nodes
(i.e. nodes with degree 1) are characterized by the smallest
diversity.  At the same time, the more densely connected groups of
nodes tended to exhibit higher diversity, with node 29 presenting the
largest diversity in this network.  Indeed, the dense connectivity of
the communities of nodes (to the left and right of node 29) allow
several self-avoiding random walks to evolve from that node.  Nodes
28, 34 and 43 exhibit the second highest diversity.  Though such facts
seem to suggest a strong correlation between diversity and node degree
(see~\cite{Costa_corrs:2007} for an investigation about the
correlation between the frequency of visits to nodes and their
degree), this is not the case.  As is clear from
Figure~\ref{fig:corrs}, which presents the scatterplot obtained by
considering the node degree and overall diversity for the network in
Figure~\ref{fig:graph}, these two measurements only a relatively weak
positive correlation can be observed between these two measurements.
Indeed, several nodes with high degree (i.e. hubs) in the GG structure
--- including nodes 11, 61, 62, 70, 73, 75, 78 --- do not have high
diversity.  On the other hand, node 45, with a high diversity, has a
very low degree (2).  Such as nodes 34 and 43, this node implements a
bridge between the two main communities in this GG structure.  At the
same time, nodes 28, 34 and 43 are hubs characterized by high
diversity.  Such a weak correlation between node degree, clear from
the scatterplot in Figure~\cite{Costa_corrs:2007}, and diversity can
be accounted by the fact that the degree is an exclusively local
property of a node, while the diversity is affected by the topological
properties of many other surrounding nodes. As a further illustration
of the more intricate nature of diversity, consider the two following
very simple networks: (i) a hub whose adjacent nodes are weakly
interconnected; and (ii) a hub with strongly interconnected adjacent
nodes.  In the former situation, after the moving agent leaves the hub
to an adjacent node $j$, it can proceed only to the few nodes
connected to $j$, implying small diversity at $h=2$ and subsequently.
Contrariwise, because of the many interconnections between the nodes
adjacent to the hub in situation (ii), many more self-avoiding walks
will be possible, increasing the diversity at $h=2$ and beyond.
Therefore, the diversity is also affected by the clustering
coefficient around each of the nodes along the self-avoiding walks.
Observe also that though most pair of adjacent nodes tend to present
similar diversities, this is not necessarily guaranteed (see, for
instance, nodes 29 and 16).  

Another interesting aspect regards the possible relationship between
diversity and community structure.  As suggested by the situation of
node 45, it could be conjectured that nodes placed between two
communities would tend to present higher diversity, as the
self-avoiding walks emanating from such nodes could proceed freely
inside both communities.  At the same time, small communities such as
that in the left-hand side of the graph in Figure~\ref{fig:graph}
would tend to have nodes exhibiting low diversity.  Because diversity
is not directly related to betweeness centrality, additional studies
are necessary in order to investigate more precisely how diversity and
communities are related.

The quantification of the overall diversity of each node allows many
interesting practical interpretations and applications.  For instance,
in case the moving agent is attacking the network, the greatest
disruption will be obtained in the cases in which it starts from high
diversity nodes, such as 29 and 43.  This would be true even if in
cases the moving agent destroys the nodes after visiting them.  At the
same time, distribution of mass of information would be most
effectively performed by allocating the sources to nodes with high
diversity.  Another interesting applications would be to consider the
spatial exploration of the network while starting from different
nodes. Exploring agents starting from low-diversity nodes will have to
invest much more efforts (i.e. steps) in order to explore the network
nodes than those starting at higher diversity nodes.  Immediate
analogies can be drawn with WWW exploration and knowledge
acquisition~\cite{Costa_know:2006}, where each node represents a piece
of knowledge which are acquired by the moving agent as it moves
through the network.  In addition, edge or node failures or attacks
taking place at low-diversity regions of the network will have higher
changes of causing major disruptions to the connectivity.

It is important to observe that experiments performed considering
diversity entropy signatures defined by traditional random walks
(i.e. with possibility of repeating edges and nodes) have led to
substantially less intuitive and informative results regarding the
structure of the analyzed networks.

\section{Concluding Remarks}

Several systems in the real-world involves non-linear transient
evolution after starting from well-defined, specific states.  Examples
of such dynamics are numerous and include WWW navigation, Internet
routing, evolution of knowledge and cultural acquisition
(e.g.~\cite{Costa_know:2006}), disease spreading, as well as
evolutionary processes leading to new species (phylogenetics), to name
but a few.  Once the underlying system has been properly represented
as a complex network, transient dynamics can be investigated by
performing diverse types of random walks.  Though several works have
been reported involving the traditional random walk, where nodes and
edges can be visited more than once, relatively few approaches have
concentrated attention in non-linear transient dynamics obtained by
performing self-avoiding random walks (e.g.~\cite{Herrero_self:2003,
Herrero_self:2005,Costa_know:2006}).  One particularly interesting
aspect of such a kind of transient dynamics concerns its necessarily
finite nature (as opposite to traditional random walks) and more
purposeful exploration and coverage of the networks, in the sense of
connecting nodes with the smallest number of edges.  Therefore, the
characterization of such a type of dynamics can provide valuable
resources not only for understanding the structure of networks, but
also for quantifying important dynamics properties such as node
coverage and redundancy/resilience.

The current work focused attention on the issue of how diverse are the
self-avoiding random walks performed in different types of networks
after starting from different nodes.  Because we wanted to obtain
information about the diversity along the several steps while moving
from a specific starting node, a simple sampling algorithm was used in
order to estimate the probabilities of visits to nodes required for
the diversity entropy calculation.  In such a way, diversity entropy
signatures can be estimated for every node in different types of
networks.  The main contributions of the current article are listed
and briefly discussed in the following:

{\bf Definition of the diversity entropy signature:} An objective way
to quantify the diversity of self-avoiding random walks has been
described which takes into account the diversity along several
subsequent steps along the walk, after starting from an individual
node $i$.  Such an approach allows the discrimination, for instance,
between walks which are diverse only at the initial steps and walks
which are diverse only at the later steps.  The estimation of the
diversity in terms of the entropy of the probabilities of visits to
nodes provides a natural and intuitive choice for measuring the
diversity.  A simple but effective sampling algorithm has been
suggested which can estimate the probability of visits to nodes at
each step along the self-avoiding walks.  The diversity entropy
signature can be used for several purposes, including the study of
network categories and the properties of nodes in specific networks.
Both these possibilities have been explored in this article.  Though
potentially affected by several other topological measurements
(e.g. node degree, clustering coefficient and betweeness centrality),
the diversity is not necessarily correlated to any of these features,
therefore providing complementary information about the topology and
dynamics of complex networks.

{\bf Characterization of the diversity of several types of networks:}
The diversity entropy signature has been used in order to investigate
the general diversity of categories of networks.  In this work, we
considered 6 diverse and representative theoretical models:
Erd\H{o}s-R\'enyi (ER), Barab\'asi-Albert (BA), Watts-Strogatz (WS)
and a geographical model (GG) --- as well as two recently introduced
knitted types of complex networks~\cite{Costa_comp:2007} --- the
path-transformed BA model (PA) and path-regular networks (PN).  The
diversity entropy signatures were estimated for all nodes in each
realization of each of these models, and summarized in terms of their
average and standard deviation for each network.  A series of
interesting results were obtained, including the verification of the
increase of diversity with the average node degree, the existence of
two types of transient dynamics (steep and gradual increase along the
steps), the high dispersion of diversity observed for the WS and GG
models, as well as the surprising uniformity of the diversity
signatures for the PN networks.  The latter model provides one of the
most effective structure ensuring steep increase of diversity for all
nodes and can be used in the design of several practical systems such
as information dissemination/exploration and transportation systems.

{\bf Definition of Adjacency as a principal cause of diversity:} The
two types of signature regimes identified among the 6 considered
theoretical models are at least partially accounted by the degree of
adjacency between any pairs of nodes in each network.  While the
standard adjacency between two nodes simply states that there is at
least one edge connecting them, the concept of \emph{degree of
adjacency} has been considered in this work in order to take into
account the adjacency implemented through longer connections.  More
specifically, the degree of adjacency between any two nodes in a
network at a given path-length is understood as the number of paths of
that length connecting the two nodes.  By using such a concept, it
becomes possible to discriminate between otherwise degree regular
models such as ER, WS, GG, PN and PA.  More specifically, because of
spatial constraints, both the WS and GG structures have most pairs of
nodes exhibiting higher degree of adjacency for several path-lengths.
Such a local connectivity enhances the chances of termination of the
self-avoiding walks, therefore reducing considerably the diversity
entropies.  Such an effect has been clearly identified in the obtained
results, leading to the identification of two regimes of transient
evolution of the diversity.

{\bf Use of sound multivariate statistics to decorrelated the
entropies:} Because the diversity entropies at subsequent steps tend
to be highly correlated, especially for more regular networks, it
becomes essential to extract the most representative information from
the signatures by decorrelating the entropies at each step.  In this
work we have applied two sound and established optimal methods for
dimensionality reduction for obtaining more representative projections
of the diversity signatures.  The \emph{canonical projection}
methodology, which performs dimensionality reduction in order to
optimally maximize the separation between the clusters produced by
each category of networks, was applied in this work in order to
emphasize the relationship between the diversity structure of the 6
considered models.  The weights assigned to each original measurement
in the case of $m=5$ by this transformation confirmed the special
information of the initial 3 or 4 diversity entropies for the
discrimination between the considered models. The \emph{principal
component analysis} approach was applied in order to decorrelate the
entropy signatures obtained for individual nodes in specific networks.
In most cases, the first principal variable was found to be mainly
defined by the initial entropies.  In the case of the WS and GG
models, the first principal variable corresponded very closely to the
arithmetic average of all entropies.  This variable has been used to
summarize the diversity of individual nodes of specific networks into
a single measurement, called \emph{overall diversity}.  Observe that
the use of the arithmetic average as a single descriptor in the case
of the WS and GG models has not been imposed a priori, but established
as a consequence of optimal decorrelation between the several
diversity entropies.  The second principal variable was found to be
more strongly affected by the first diversity entropies, confirming
the importance of these features.

{\bf Characterization of the diversity of several types of networks at
the individual node level:} The diversity entropy signature was also
investigated at the level of individual node for an example network
from each of the 6 considered types of structures.  Such an analysis
led to results similar to those obtained for the global analysis.  By
using the principal variable as a single quantification of the
diversity of the individual nodes, we were able to study in more
detail a sample of GG network, which provides the spatial position of
the nodes and exhibit community structure.  The overall diversity was
estimated for each node and 9 intervals of diversity were defined by
binning the exponential values of the overall diversity, which allowed
a more uniform distribution of the measurements.  The nodes of the
network were then identified (colored) according to such intervals,
leading to an interesting partitioning of the network in terms of the
respective diversities.  A series of interesting results were
obtained.  First, the low-diversity nodes tended to appear at the
borders of the network, often involving an extremity node.
Interestingly, the own concept of diversity can be used to define the
network border nodes as corresponding to those which have little
access to the remaining network.  Second, the more densely
interconnected nodes tended to present high diversity.  However, no
strong correlation has been identified, at least for the considered GG
network, between node degree and diversity. Possible relationships
between diversity and community structure have also been identified.

Such diverse results are not interesting by themselves regarding
several aspects of complex network research, but also open several
possibilities for future exploration, including but not being limited
to: (i) study of other types of walks, such as preferential; (ii)
applications to the characterization of real-world networks; (iii)
study of scaling effects, especially with $N$; (iv) consider the
length of the self-avoiding walks as complementary information about
the structure of the networks; (v) investigate more systematically the
relationship between the degree of adjacency and the diversity,
especially the possibility if the latter can be predicted by the
former; (vi) devise network growth algorithms based on diversity
constraints; (vii) use of the diversity to identify particularly weak
and stronger points in networks (e.g. with respect to resilience or
distribution) and try to enhance such situations; and (viii)
investigate further the possible relationship between diversity and
communities, especially regarding possible influences between
diversity and betweeness centrality.

\begin{acknowledgments}
Luciano da F. Costa thanks CNPq (308231/03-1) and FAPESP (05/00587-5)
for sponsorship.
\end{acknowledgments}

\bibliography{diverse}

\begin{thebibliography}{23}
\expandafter\ifx\csname natexlab\endcsname\relax\def\natexlab#1{#1}\fi
\expandafter\ifx\csname bibnamefont\endcsname\relax
  \def\bibnamefont#1{#1}\fi
\expandafter\ifx\csname bibfnamefont\endcsname\relax
  \def\bibfnamefont#1{#1}\fi
\expandafter\ifx\csname citenamefont\endcsname\relax
  \def\citenamefont#1{#1}\fi
\expandafter\ifx\csname url\endcsname\relax
  \def\url#1{\texttt{#1}}\fi
\expandafter\ifx\csname urlprefix\endcsname\relax\def\urlprefix{URL }\fi
\providecommand{\bibinfo}[2]{#2}
\providecommand{\eprint}[2][]{\url{#2}}

\bibitem[{\citenamefont{Albert and Barab\'asi}(2002)}]{Albert_Barab:2002}
\bibinfo{author}{\bibfnamefont{R.}~\bibnamefont{Albert}} \bibnamefont{and}
  \bibinfo{author}{\bibfnamefont{A.~L.} \bibnamefont{Barab\'asi}},
  \bibinfo{journal}{Rev. Mod. Phys.} \textbf{\bibinfo{volume}{74}},
  \bibinfo{pages}{47} (\bibinfo{year}{2002}).

\bibitem[{\citenamefont{Dorogovtsev and Mendes}(2002)}]{Dorogov_Mendes:2002}
\bibinfo{author}{\bibfnamefont{S.~N.} \bibnamefont{Dorogovtsev}}
  \bibnamefont{and} \bibinfo{author}{\bibfnamefont{J.~F.~F.}
  \bibnamefont{Mendes}}, \bibinfo{journal}{Advs. in Phys.}
  \textbf{\bibinfo{volume}{51}}, \bibinfo{pages}{1079} (\bibinfo{year}{2002}).

\bibitem[{\citenamefont{Newman}(2003)}]{Newman:2003}
\bibinfo{author}{\bibfnamefont{M.~E.~J.} \bibnamefont{Newman}},
  \bibinfo{journal}{SIAM Rev.} \textbf{\bibinfo{volume}{45}},
  \bibinfo{pages}{167} (\bibinfo{year}{2003}).

\bibitem[{\citenamefont{Boccaletti et~al.}(2006)\citenamefont{Boccaletti,
  Latora, Moreno, Chavez, and Hwang}}]{Boccaletti:2006}
\bibinfo{author}{\bibfnamefont{S.}~\bibnamefont{Boccaletti}},
  \bibinfo{author}{\bibfnamefont{V.}~\bibnamefont{Latora}},
  \bibinfo{author}{\bibfnamefont{Y.}~\bibnamefont{Moreno}},
  \bibinfo{author}{\bibfnamefont{M.}~\bibnamefont{Chavez}}, \bibnamefont{and}
  \bibinfo{author}{\bibfnamefont{D.}~\bibnamefont{Hwang}},
  \bibinfo{journal}{Phys. Rep.} \textbf{\bibinfo{volume}{424}},
  \bibinfo{pages}{175} (\bibinfo{year}{2006}).

\bibitem[{\citenamefont{da~F.~Costa
  et~al.}(2007{\natexlab{a}})\citenamefont{da~F.~Costa, Rodrigues, Travieso,
  and Boas}}]{Costa_surv:2007}
\bibinfo{author}{\bibfnamefont{L.}~\bibnamefont{da~F.~Costa}},
  \bibinfo{author}{\bibfnamefont{F.~A.} \bibnamefont{Rodrigues}},
  \bibinfo{author}{\bibfnamefont{G.}~\bibnamefont{Travieso}}, \bibnamefont{and}
  \bibinfo{author}{\bibfnamefont{P.~R.~V.} \bibnamefont{Boas}},
  \bibinfo{journal}{Advs. in Phys.} \textbf{\bibinfo{volume}{56}},
  \bibinfo{pages}{167} (\bibinfo{year}{2007}{\natexlab{a}}).

\bibitem[{\citenamefont{Doyle and Snell}(1984)}]{Doyle_Snell:1984}
\bibinfo{author}{\bibfnamefont{P.~G.} \bibnamefont{Doyle}} \bibnamefont{and}
  \bibinfo{author}{\bibfnamefont{J.~L.} \bibnamefont{Snell}},
  \emph{\bibinfo{title}{Random Walks and electric networks}}
  (\bibinfo{publisher}{Carus Mathematical Monographs}, \bibinfo{year}{1984}).

\bibitem[{\citenamefont{Sethna}(2006)}]{Sethna:2006}
\bibinfo{author}{\bibfnamefont{J.~P.} \bibnamefont{Sethna}},
  \emph{\bibinfo{title}{Entropy, order parameters, and complexity}}
  (\bibinfo{publisher}{Oxford University Press}, \bibinfo{year}{2006}).

\bibitem[{\citenamefont{da~F.~Costa
  et~al.}(2007{\natexlab{b}})\citenamefont{da~F.~Costa, Sporns, Antiqueira,
  Nunes, and Oliveira}}]{Costa_corrs:2007}
\bibinfo{author}{\bibfnamefont{L.}~\bibnamefont{da~F.~Costa}},
  \bibinfo{author}{\bibfnamefont{O.}~\bibnamefont{Sporns}},
  \bibinfo{author}{\bibfnamefont{L.}~\bibnamefont{Antiqueira}},
  \bibinfo{author}{\bibfnamefont{M.~G.~V.} \bibnamefont{Nunes}},
  \bibnamefont{and} \bibinfo{author}{\bibfnamefont{O.~N.}
  \bibnamefont{Oliveira}}, \bibinfo{journal}{Appl. Phys. Letts.}
  \textbf{\bibinfo{volume}{91}}, \bibinfo{pages}{054107}
  (\bibinfo{year}{2007}{\natexlab{b}}).

\bibitem[{\citenamefont{Noh and Rieger}(2004)}]{Rieger:2004}
\bibinfo{author}{\bibfnamefont{J.~D.} \bibnamefont{Noh}} \bibnamefont{and}
  \bibinfo{author}{\bibfnamefont{H.}~\bibnamefont{Rieger}},
  \bibinfo{journal}{Phys. Rev. Letts.} \textbf{\bibinfo{volume}{92}},
  \bibinfo{pages}{118701} (\bibinfo{year}{2004}),
  \bibinfo{note}{arXiv:cond-mat/0307719}.

\bibitem[{\citenamefont{Masuda and Konno}(2004)}]{Masuda:2004}
\bibinfo{author}{\bibfnamefont{N.}~\bibnamefont{Masuda}} \bibnamefont{and}
  \bibinfo{author}{\bibfnamefont{N.}~\bibnamefont{Konno}},
  \bibinfo{journal}{Phys. Rev. E} \textbf{\bibinfo{volume}{69}},
  \bibinfo{pages}{066113} (\bibinfo{year}{2004}),
  \bibinfo{note}{arXiv:cond-mat/0401255}.

\bibitem[{\citenamefont{Pons and Latapy}(2005)}]{Pons_comm:2005}
\bibinfo{author}{\bibfnamefont{P.}~\bibnamefont{Pons}} \bibnamefont{and}
  \bibinfo{author}{\bibfnamefont{M.}~\bibnamefont{Latapy}}
  (\bibinfo{year}{2005}), \bibinfo{note}{arXiv:physics/0512106}.

\bibitem[{\citenamefont{Eisler and Kertesz}(2005)}]{Eisler:2005}
\bibinfo{author}{\bibfnamefont{Z.}~\bibnamefont{Eisler}} \bibnamefont{and}
  \bibinfo{author}{\bibfnamefont{J.}~\bibnamefont{Kertesz}},
  \bibinfo{journal}{Phys. Rev. E} \textbf{\bibinfo{volume}{71}},
  \bibinfo{pages}{057104} (\bibinfo{year}{2005}),
  \bibinfo{note}{arXiv:physics/0512106}.

\bibitem[{\citenamefont{Zhou}(2003)}]{Zhou:2003}
\bibinfo{author}{\bibfnamefont{H.}~\bibnamefont{Zhou}}, \bibinfo{journal}{Phys.
  Rev. E} \textbf{\bibinfo{volume}{67}}, \bibinfo{pages}{061901}
  (\bibinfo{year}{2003}), \bibinfo{note}{arXiv:physics/0302032}.

\bibitem[{\citenamefont{Kinouchi et~al.}(2002)\citenamefont{Kinouchi, Martinez,
  Lima, Lourenco, and Risau-Gusman}}]{Kinouchi_thesaurus:2001}
\bibinfo{author}{\bibfnamefont{O.}~\bibnamefont{Kinouchi}},
  \bibinfo{author}{\bibfnamefont{A.~S.} \bibnamefont{Martinez}},
  \bibinfo{author}{\bibfnamefont{G.~F.} \bibnamefont{Lima}},
  \bibinfo{author}{\bibfnamefont{G.~M.} \bibnamefont{Lourenco}},
  \bibnamefont{and}
  \bibinfo{author}{\bibfnamefont{S.}~\bibnamefont{Risau-Gusman}},
  \bibinfo{journal}{Physica A} \textbf{\bibinfo{volume}{315}},
  \bibinfo{pages}{665} (\bibinfo{year}{2002}).

\bibitem[{\citenamefont{Yang}(2005)}]{Yang:2005}
\bibinfo{author}{\bibfnamefont{S.~J.} \bibnamefont{Yang}},
  \bibinfo{journal}{Phys. Rev. E} \textbf{\bibinfo{volume}{71}},
  \bibinfo{pages}{016107} (\bibinfo{year}{2005}).

\bibitem[{\citenamefont{da~F.~Costa}(2006)}]{Costa_know:2006}
\bibinfo{author}{\bibfnamefont{L.}~\bibnamefont{da~F.~Costa}},
  \bibinfo{journal}{Phys. Rev. E} \textbf{\bibinfo{volume}{74}},
  \bibinfo{pages}{026103} (\bibinfo{year}{2006}).

\bibitem[{\citenamefont{da~F.~Costa}(2007{\natexlab{a}})}]{Costa_path:2007}
\bibinfo{author}{\bibfnamefont{L.}~\bibnamefont{da~F.~Costa}}
  (\bibinfo{year}{2007}{\natexlab{a}}), \bibinfo{note}{arXiv:0711.1271}.

\bibitem[{\citenamefont{da~F.~Costa}(2007{\natexlab{b}})}]{Costa_comp:2007}
\bibinfo{author}{\bibfnamefont{L.}~\bibnamefont{da~F.~Costa}}
  (\bibinfo{year}{2007}{\natexlab{b}}), \bibinfo{note}{arXiv:0711.2736}.

\bibitem[{\citenamefont{da~F.~Costa}(2007{\natexlab{c}})}]{Costa_longest:2007}
\bibinfo{author}{\bibfnamefont{L.}~\bibnamefont{da~F.~Costa}}
  (\bibinfo{year}{2007}{\natexlab{c}}), \bibinfo{note}{arXiv:0712.0415}.

\bibitem[{\citenamefont{Herrero and Saboya}(2005)}]{Herrero_self:2005}
\bibinfo{author}{\bibfnamefont{C.~P.} \bibnamefont{Herrero}} \bibnamefont{and}
  \bibinfo{author}{\bibfnamefont{M.}~\bibnamefont{Saboya}},
  \bibinfo{journal}{Phys. Rev. E} \textbf{\bibinfo{volume}{71}},
  \bibinfo{pages}{016103} (\bibinfo{year}{2005}).

\bibitem[{\citenamefont{Herrero}(2003)}]{Herrero_self:2003}
\bibinfo{author}{\bibfnamefont{C.~P.} \bibnamefont{Herrero}},
  \bibinfo{journal}{Phys. Rev. E} \textbf{\bibinfo{volume}{68}},
  \bibinfo{pages}{026106} (\bibinfo{year}{2003}).

\bibitem[{\citenamefont{da~F.~Costa and Cesar}(2001)}]{Costa_book:2001}
\bibinfo{author}{\bibfnamefont{L.}~\bibnamefont{da~F.~Costa}} \bibnamefont{and}
  \bibinfo{author}{\bibfnamefont{R.~M.} \bibnamefont{Cesar}},
  \emph{\bibinfo{title}{Shape Analysis and Classification: {T}heory and
  Practice}} (\bibinfo{publisher}{CRC Press}, \bibinfo{year}{2001}).

\bibitem[{\citenamefont{McLachlan}(1998)}]{McLachlan:1998}
\bibinfo{author}{\bibfnamefont{G.~J.} \bibnamefont{McLachlan}},
  \emph{\bibinfo{title}{Discriminant Analysis and Statistical Pattern
  Recognition}} (\bibinfo{publisher}{John Wiley and Sons},
  \bibinfo{year}{1998}).

\end{thebibliography}
\end{document}